\documentclass[sigconf]{acmart}

\usepackage{algorithm}
\usepackage{algorithmicx}
\usepackage{algpseudocode}
\usepackage{hyperref}
\usepackage{subcaption}
\usepackage{setspace}
\usepackage{url}
\usepackage{amsmath}
\usepackage{amsthm}
\usepackage{tikz}
\usepackage{graphicx}
\usepackage{proof}
\usepackage{enumitem}
\usepackage{forest}
\usepackage{bbm}
\usepackage{mathrsfs}
\usepackage{bookmark}
\usepackage{caption}
\usepackage{float}
\usepackage{tabularx}
\usepackage{booktabs}
\usepackage{multirow}
\usepackage{tcolorbox}

\usetikzlibrary{arrows.meta,shapes,positioning,shadows,trees}

\tikzset{
    basic/.style  = {draw, text width=2cm, drop shadow, font=\sffamily, rectangle},
    root/.style   = {basic, rounded corners=2pt, thin, align=center,
                     fill=green!30},
    onode/.style = {basic, thin, rounded corners=2pt, align=center, fill=green!60,text width=3cm,},
    tnode/.style = {basic, thin, align=left, fill=pink!60, text width=6.5em},
    edge from parent/.style={draw=black, edge from parent fork right}
}

\floatstyle{ruled}
\newfloat{problem}{htbp}{lop}
\floatname{problem}{Problem}

\newtheorem{theorem}{Theorem}

\newtheorem{conjecture}[theorem]{Conjecture}

\theoremstyle{definition}
\newtheorem{question}{Question}

\DeclareMathOperator{\EX}{\mathbb{E}}
\DeclareMathOperator{\Ind}{\mathbbm{1}}

\acmYear{2020}\copyrightyear{2020}
\setcopyright{acmlicensed}
\acmConference[AFT '20]{2nd ACM Conference on Advances in Financial Technologies}{October 21--23, 2020}{New York, NY, USA}
\acmBooktitle{2nd ACM Conference on Advances in Financial Technologies (AFT '20), October 21--23, 2020, New York, NY, USA}
\acmPrice{15.00}
\acmDOI{10.1145/3419614.3423261}
\acmISBN{978-1-4503-8139-0/20/10}

\begin{document}

\title{Stablecoins 2.0: Economic Foundations and Risk-based Models}


\author{Ariah Klages-Mundt}
\affiliation{\institution{Cornell University}}

\author{Dominik Harz}
\affiliation{\institution{Imperial College London}}

\author{Lewis Gudgeon}
\affiliation{\institution{Imperial College London}}

\author{Jun-You Liu}
\affiliation{\institution{Cornell University}}

\author{Andreea Minca}
\affiliation{\institution{Cornell University}}

\renewcommand{\shortauthors}{Klages-Mundt, et al.}


\begin{abstract}
Stablecoins are one of the most widely capitalized type of cryptocurrency.
However, their risks vary significantly according to their design and are often poorly understood.
We seek to provide a sound foundation for stablecoin theory, with a risk-based functional characterization of the economic structure of stablecoins.
First, we match existing economic models to the disparate set of custodial systems.
Next, we characterize the unique risks that emerge in non-custodial stablecoins and develop a model framework that unifies existing models from economics and computer science.
We further discuss how this modeling framework is applicable to a wide array of cryptoeconomic systems, including cross-chain protocols, collateralized lending, and decentralized exchanges.
These unique risks yield unanswered research questions that will form the crux of research in decentralized finance going forward.

\end{abstract}



\keywords{Stablecoins, Risk, Governance, Capital Structure Models, DeFi}


\maketitle
\pagestyle{plain} 

\section{Introduction}

Stablecoins are cryptocurrencies with an added economic structure that aims to stabilize their price and purchasing power.
There are two classes of stablecoin: custodial, which require trust in a third party, and non-custodial, which replace this trust with economic mechanisms.
Major custodial examples such as Tether, Binance USD, USDC, and TrueUSD have a combined market capitalization of over USD 10bn. 
On the non-custodial side, of the USD 1bn of value locked in so-called Decentralized Finance (DeFi) protocols, more than 50\% are allocated to Maker's Dai stablecoin.

Several recent papers and industry reports provide overviews of stablecoins \cite{bullmann2019,moin2019classification,pernice2019,mita2019,blockchain2019,samman2019}. 
These typically categorize stablecoins based on the type of collateral used, peg target, and technological mechanics (e.g., on-chain, off-chain, algorithmic) and informally relate stablecoin mechanisms to traditional monetary tools (e.g., interest rates). 
The history of money and stablecoins, and the institutional structures of stablecoins are discussed in~\cite{lipton2020_10}. 
The regulatory perspective of stablecoins, including classification, regulatory gaps, and systemic stability risks are discussed in~\cite{adachi2020}.

In this paper our fundamental aim is different.
Market events have demonstrated that even stablecoins---supposedly price stable---can exhibit significant volatility. 
On the 12th March 2020, amidst the SARS-COV-2 pandemic, market volatility affected the stablecoin Dai~\cite{makerdao} so severely that it entered a deflationary deleveraging spiral, forcing it to deviate from its peg.
While the aforementioned papers observe and categorize \emph{existing} stablecoin designs, none of the works develop risk-based models of a broad design space of \emph{possible} choices and their fundamental trade-offs.
Here we seek to fill this gap, providing sound economic foundations to inform stablecoin design, focusing on financial risk.
As such, the work is intended to serve as a ``manual" for future stablecoin research.

Firstly, we provide an overview of the relevant risk-based models from economics and computer science, seeking to avoid duplication of work by only extending models where necessary.
Secondly, we provide a number of formalized open questions drawing on capital structure theory.
Throughout we assume that stablecoin systems are used and operated by economically rational agents whose actions ultimately determine the stability and security of these systems.
However, we do not solve the stated open problems in the context of this paper.
This work builds on the previous attacks on decentralized stablecoins identified in~\cite{klagesmundt2019}.


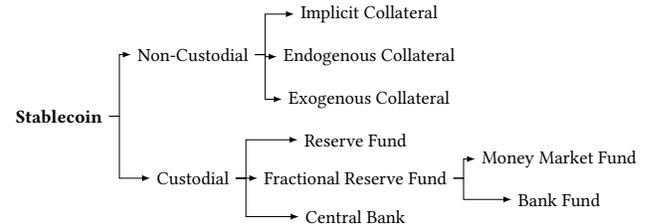
\begin{figure}
\resizebox{1\columnwidth}{!}{%
\begin{forest} for tree={
    grow=east,
    growth parent anchor=east,
    parent anchor=east,
    child anchor=west,
    edge path={\noexpand\path[\forestoption{edge},->, >={latex}] 
         (!u.parent anchor) -- +(5pt,0pt) |- (.child anchor)
         \forestoption{edge label};}
}
          [\textbf{Stablecoin}
    [Custodial
		[Central Bank]
        [Fractional Reserve Fund
            [Bank Fund]
            [Money Market Fund]
        ]
        [Reserve Fund]
    ]
    [Non-Custodial
        [Exogenous Collateral]
        [Endogenous Collateral]
        [Implicit Collateral
        ]
    ]
]
\end{forest}
}
\caption{Risk-based overview of stablecoin design space.}
\label{fig:categories-forest}
\end{figure}

We uncover five central dimensions of risks. 
In non-custodial stablecoins: (1) effects from deleveraging-like processes on collateral-like assets and risk in underlying collateral-like thing (as discussed, e.g., in \cite{klagesmundt2019,klagesmundt2020}), (2) data feed and governance risks, (3) base layer risks from mining incentives, and (4) smart contract coding risks, on which the formal verification literature can be applied. 
In contrast, in custodial stablecoins, the first applies in a very different way to affect issuer incentives as well as an additional central risk dimension of (5) censorship and counterparty risk. 
Our stablecoin mechanism categorization decomposes the design space according to these dimensions of risk.
Figure~\ref{fig:categories-forest} summarizes our categorization along some of the most important dimensions of risk.

\subsection*{Contributions}


\begin{itemize}
    \item 
    We provide a functional breakdown of custodial stablecoin designs with a correspondence to taxonomy and models for traditional financial instruments (Section~\ref{sec:custodial}).
    \item 
    We provide a common functional framework for relating the economic mechanics of all non-custodial stablecoin designs and a discussion of new risks that emerge in this setting (Section~\ref{sec:noncustodial}).
    \item 
    We provide questions of economic stability and security that apply in evaluating non-custodial stablecoins (Section~\ref{sec:noncustodial}).
    \item 
    We provide a framework of models toward measuring stability and security including open research questions based on agents' decisions (Section~\ref{sec:questions}).
    \item 
    We provide methods for estimating agents' preferences as represented by utility functions, providing a minimal working example using historical data from Maker (Section~\ref{sec:questions}).
    \item
    Last, we outline how our models can be applied to DeFi protocols including composite stablecoins, cross-chain and syntehtic assets as well as lending protocols and decentralized exchanges (Section~\ref{sec:discussion}).
\end{itemize}






\section{Custodial Stablecoins}
\label{sec:custodial}

In custodial stablecoins, custodians are entrusted with off-chain collateral assets, such as fiat currencies, bonds, or commodities.
An issuer (possibly the same entity) then offers digital tokens to represent an on-chain version of a reserve asset (e.g., USD). 
Holders of the digital token have some form of claim against the custodial assets, which maintains the peg. 
The custodial assets include \emph{reserve assets}, which are what the stablecoin is pegged against (e.g., USD), and \emph{capital assets}, which are other assets that back stablecoin supply. 
Capital assets are comparable to illiquid assets held by a bank and short-term treasuries held by money market funds.

Custodial stablecoins introduce coin holders to \textit{counterparty} and \textit{censorship} risks related to the off-chain assets and \emph{economic} risks of the capital assets. 
These risks are similar to risks in traditional assets. 
Counterparty risks may be heightened due to the shared account structure with the custodian and lack of government deposit insurance. 
In the event that the central entities are unable to fulfill their obligations (e.g., the result of fraud, mismanagement, theft, or government seizure), the stablecoin value can go to zero.
Table~\ref{table:custodial_category} summarizes categories, applicable models, and projects.

\subsection{Reserve Fund = 100\% reserve off-chain}

In Reserve Fund stablecoins, the stablecoin maintains a 100\% reserve ratio--i.e., each stablecoin is backed by a unit of the reserve asset (e.g., 1 USD) held by the custodian. The price target is maintained via two mechanisms. 
Coins may be directly redeemable off-chain for the underlying reserve asset. 
In this case, arbitrage trades incentivize external actors to close any price deviations that occur. 
Alternatively, the issuer may designate `authorized participants' (possibly the issuer itself) who alone have the ability to create and redeem stablecoins against the reserve. 
In this case, the authorized participants capture price deviation arbitrage.


Reserve Fund stablecoins resemble the structures of e-money, narrow banks, and currency boards. 
E-money is a prepaid bearer instrument. 
Deposits at a narrow bank are backed by 100\% reserves held at a central bank. 
A currency board maintains a fixed exchange rate of a sovereign currency using 100\% reserves in a foreign currency (e.g., the Hong Kong Dollar maintains a USD peg using USD reserves). 
Of these, the Reserve Fund stablecoin most closely mirrors the currency board as the market price of the stablecoin floats subject to creation and redemption similarly to how the sovereign currency floats subject to creation and redemption of the currency board. 
On the other hand, e-money and narrow bank deposits are treated identically with the currency itself. 
Notably, unlike the currency board, the stablecoin reserves may be stored in commercial bank deposit accounts, which may bear bank run risks.
We discuss approaches to modeling Reserve Fund stablecoins in Appendix~\ref{sec:appendix-reservefund}.

\subsection{Fractional Reserve Fund}
A Fractional Reserve Fund stablecoin is backed by a mixture of reserve assets and other capital assets, and has a target price. 
The fund holds reserves in a target asset (or other highly liquid stable assets) that account for $<100\%$ of the stablecoin supply in order to facilitate stablecoin redemptions. 
Similar to the Reserve Fund design, these reserve assets may resemble commercial bank deposits which exceed the government deposit insurance level, in which case they may take on commercial bank run risk. 
The other capital assets account for the remaining stablecoin supply value and earn a higher interest rate for the stablecoin issuer. 
The capital assets can be liquidated to handle additional stablecoin redemptions, but are subject to price risk. 
Within this class, the important dividing point is the type of capital assets held: illiquid assets (similar to a commercial bank) or low-risk assets (similar to a money market fund). 
In either case, the stablecoin has a floating price, and so the peg is maintained through similar ETF arbitrage trades involving fund redemptions. 
Thus applicable risk models would take the form of ETF models in serial with bank run or money market models, which we discuss next.
We provide further detail on each type of stablecoin in Appendix~\ref{sec:appendix-fractionalreservefund}.


\subsection{Central Bank Digital Currency}

Central Bank Digital Currency (CBDC) is a consumer-facing fiat digital currency that aims to provide a risk-free store of value. 
CBDC proposes a different monetary system to the status quo. 
Currently, central bank reserve deposits are available to commercial banks, but not to consumers or non-bank businesses. 
Consumers and businesses hold commercial bank accounts. 
The non-cash money supply is determined by the lending of commercial banks (see \cite{mcleay2014}). 
The government intervenes in this monetary system to create risk-free consumer deposit accounts by providing commercial bank deposit insurance. 
Instead, CBDC provides consumer-facing deposits at the central bank.\footnote{See \cite{auer2020} for a discussion on design and architecture of CBDC. 
The version comparable to stablecoins is the token-based design.}

CBDC represents a change in the structure of money deposits within the banking system and not a change in the currency stability model itself. 
In fact, CBDC is in many ways a more ideal setting for existing currency models as it is closer in form to fiat than commercial bank deposits. 
Traditional currency models like \cite{morris1998} and \cite{guimaraes2007} apply to understand the stability of fiat currencies. 
These models typically assume that the central bank/government is stability-seeking for its own sake as opposed to private banks discussed above, which are profit-seeking. 
A fiat currency is assumed to have the backing of a given country's economy, which provides a natural demand from economic activity in the currency, as well as military power and legal system. 
Given this setting, agents in these models hedge their current positions to account for demand in a next period, some of which occurs in the fiat currency and other of which occurs in a foreign currency, under a potential currency attack from an attacking agent. 
The ability to maintain a peg in this setting will depend on a relationship between reserves held by the central bank and economic demand. 

Research questions around CBDC focus on wider economic effects and indirect effects on stability, such as through commercial bank lending, credit availability, and funding in the real economy. 
\cite{barrdear2016} models the effects of CBDC on the wider economy through competition with commercial bank deposits. 
\cite{ney2019} explores the effect of CBDC on commercial bank lending to the real economy through a case study analysis of government subsidies.

\section{Non-custodial Stablecoins}
\label{sec:noncustodial}

Non-custodial stablecoins aim to be independent of the societal institutions that custodial designs rely on. They achieve this by establishing economic structure between participants implemented through smart contracts. In this setting, directly confiscating assets is prevented by the underlying blockchain mechanism.

Non-custodial stablecoins structurally resemble dynamic versions of risk transfer instruments, such as collateralized debt obligations (CDO) and contracts for difference (CFD).\footnote{They also resemble perpetual swaps, which are relatively new products on cryptocurrency exchanges.} CDOs are backed by a pool of collateral assets and sliced into tranches. Any losses are absorbed first by the junior tranche; a senior tranche only absorbs losses if the junior tranche is wiped out.

Functionally, a non-custodial stablecoin system contains the following components in some form:
\begin{itemize}
	\item \emph{Primary value}: the economic structure of the base value in the stablecoin. This is an abstracted concept of collateral with the following types: \emph{exogenous} when the collateral has primary outside use cases, \emph{endogenous} when the collateral is created for the purpose of being collateral, and \emph{implicit} when the design lacks explicit collateralization.
	
	\item \emph{Risk absorbers}: speculative agents who absorb risk and profit in the system ($\sim$ the junior tranche of a CDO).
	
	\item \emph{Stablecoin holders}: agents who make up the demand side of the stablecoin market ($\sim$ senior tranche holder of a CDO).
	
	\item \emph{Issuance}: a function performed by an agent or algorithm that determines stablecoin issuance ($\sim$ how levered a CDO is), including a deleveraging process to reduce stablecoin supply.
	
	\item \emph{Governance}: a function performed by an agent or algorithm to manage system parameters, such as deleveraging factors and price feeds, and collects a fee on system operation ($\sim$ an equity position in managing CDOs).
	
	\item \emph{Data feed}: a function to import external asset data (e.g., exchange price of assets in USD) into the blockchain virtual machine so that it is readable by the system's smart contracts.
	
	\item \emph{Miners}: agents who decide the inclusion and ordering of actions in the base blockchain layer (PoW or PoS).
\end{itemize}
The specific form of components may differ, but the general functions are universal across stablecoin designs. Depending on the design, several functions may be performed by a single agent type and others may be algorithmic. Notice that the last three components can be simplified out of traditional financial models because of legal protections; in traditional systems, we typically assume these processes are mechanical as opposed to strategic actions. As a result, stablecoins are susceptible to new manipulation attacks around governance, price feeds, and miner-extractable value (MEV).

\paragraph{Analogy to traditional monetary system}
We provide an illustration between the Maker stablecoin system\footnote{The most capitalized non-custodial stablecoin system as of 10 June 2020.} and the traditional monetary system to aid the reader in understanding the components and functional differences. In Maker, \emph{vaults} absorb risk and perform issuance. Vaults deposit ETH collateral (primary value), issue Dai secured against this collateral, and invest proceeds from Dai issuance to achieve a leveraged position. The fiat system contains a central bank, commercial bank, and depositors. The central bank regulates commercial banks and holds bank currency reserves. Commercial banks decide the money supply through lending. Depositors hold fiat currency accounts at commercial banks.

Maker vaults are parallel to commercial banks in that they both they decide money supply based on issuance incentives. For banks, this depends on profitability of lending, which incorporates the spread between long-term and short-term rates, subject to balance sheet and regulatory constraints and depositor withdrawal expectations. Vaults make a different bet collateral leverage.\footnote{Commercial bank money supply is often described as a `money multiplier' based on the required reserve ratio. This is only accurate if we assume that banks lend the maximum allowed by their constraints. This need not be the case that the optimal lending always has a binding constraint. Similarly, vaults in Maker typically do not issue stablecoins to the maximum extent of the collateral factor.} Governance is parallel to the central bank. The central bank sets rates to target economic stability and capital requirements for banks. Models typically assume the central bank mechanically targets stability by mandate. Stablecoin governance takes a different form. Governance sets rates and collateral factors to maximize system profits, which we hope to be aligned with stability. Stablecoin holders are parallel to depositors. Whereas bank depositors are guaranteed deposit redemption, stablecoin holders may have no such guarantee. Instead, they must hope that system incentives are aligned to make the stablecoin floating price stable and liquid.

A final useful parallel is in governance attacks. Through setting system parameters, stablecoin governors could inherently steal the value locked in the system, something we discuss in the context of models in the next section. A parallel attack in the traditional monetary system would be an infinite printing of money by the central bank, to the benefit of the government.

\subsection{Primary Value}

The primary value is an abstract concept of collateral that is the basis for value in the stablecoin system. It incorporates the value of collateral with explicit market prices and/or non-tokenized value  `in the system' coordinated among participants, which we term \emph{implicit collateral}. This primary value is derived from market expectations in some system. For exogenous cryptocurrency collateral (e.g., ETH), this is expectations and `confidence' about Ethereum. In implicit collateral, it is coordinated `confidence' in the stablecoin system itself. In comparison, in fiat currencies, this is confidence in a nation's government, economy, and legal system. In gold-backed currencies, it is confidence in gold.\footnote{At some level, confidence in \emph{something} seems unavoidable as a source of value in a monetary system.} In tokenized assets, it may be confidence in the custodian and expectations about cashflows of the underlying assets.

\paragraph{Exogenous collateral}
An exogenous collateral is an asset that has uses outside of the stablecoin system and for which only a small portion may be tied up in collateral for the stablecoin.
An example is ETH in Maker.
Stablecoins are issued against this collateral subject to a collateral factor that dictates the minimum over-collateralization allowed in the system. From a model perspective, the prices of exogenous collateral can be modeled exogenously.

\paragraph{Endogenous collateral}
An endogenous collateral is an asset created with the purpose of being collateral for the stablecoin. This means that it has few, if any, competing uses outside of the stablecoin system.
Examples include SNX in Synthetix (in which issuance is agent-based) and `shares' in seigniorage shares (in which issuance is algorithmic) \cite{Sams2015}).
In seigniorage shares, an `equity'-like position insures the system against price risk, absorbing losses when stablecoin demand is low and the supply needs to be contracted, and receiving newly minted stablecoins when demand is high and the supply needs to be expanded.
\footnote{While, in general, seigniorage shares has a risk absorbing effect, extremes of the idea (Ampleforth) are really just a twist on a fixed supply cryptocurrency misinterpreted as a stablecoin. Ampleforth transforms price volatility into supply volatility (e.g., daily stock splits) without having an \textit{economically} stabilizing effect on purchasing power (though may have a \textit{psychological} effect). Thus it can be interpreted as akin to seigniorage shares where all positions are the `shares' and so in fact no positions are stabilized.}
The price of endogenous collateral cannot be modeled exogenously due to endogenous feedback effects between stablecoin usage and collateral value. Its value is derived from a self-fulfilling coordination of `confidence' between its participants. 

For instance, in a crisis of confidence, if expectations of stablecoin holder demand are low, then the value of the endogenous collateral should be low, which will further shake confidence in the system and demand. On the other hand, high expectations can be self-fulfilling: with high collateral value, the stablecoin is, in a sense, more secure. If stablecoin holder demand is high, then a high price of the endogenous collateral can be justified. 

The distinction between exogenous and endogenous collateral may be best conceptualized as a spectrum. 
For instance, selected collateral has outside uses but are significantly intertwined with the stablecoin (e.g., Steem Dollars) and some stablecoins are backed by a collateral basket, including both exogenous and endogenous collateral (e.g., Celo). From a model perspective, this spectrum can be represented as the strength of these feedback effects.

\paragraph{Implicit collateral}
Some stablecoin designs do not have explicit collateral but instead propose market mechanisms to dynamically adjust supply to stabilize price. These designs work when speculators can be incentivized to absorb losses when the supply needs to be decreased by the prospect for rewards when the stablecoin supply needs to increase. We draw a parallel between the positions of such speculators and the endogenous collateral case with important functional differences. Both obtain value from self-fulfilling coordination of confidence in the stablecoin from usage and speculative expectations between the participants. Endogenous collateral represents the explicit tokenization of this, including obligation to absorb losses during supply decreases, which means it has a directly observable market price. Implicit collateral is not explicitly tokenized \emph{and} risk absorbers do not have direct obligations to absorb losses. For modeling, implicit collateral can be interpreted like endogenous collateral behind-the-scenes and accounting for this difference in financial structure of risk absorbers. The behind-the-scenes `market price' of this coordination will only be indirectly observable in the levels of stablecoin and speculative demand. However, they will play a similar role to endogenous collateral in valuing both the speculative and stablecoin positions. The stability of both endogenous and implicit collateral stablecoins will rely on how participants perceive and coordinate this value over time.

One type includes Basis \cite{Basis2018} and NuBits \cite{Nubits2014}. In these designs `shares' are awarded if stablecoin supply increases, but do not necessarily face direct losses when supply contracts (but, of course, they do face indirect losses from the share market price). Supply contraction relies on selling `bond' positions to remove stablecoins from circulation in return for future rewards when supply is next increased. In Basis, this is algorithmic, whereas in NuBits, this is coordinated through share voting (and a couple other stabilization mechanisms, including share demurrage, are available for voters to choose from). 
If we tokenize an obligation to purchase `bonds' during contractions and combine with `shares' positions, then the result resembles seigniorage shares. As it is not tokenized in this way, the equivalent of `collateral' is only implicit with no observable market price. Comparatively, seigniorage `shares' ought to be valued differently to be compensated for extra obligation. And downside price stabilization will depend on incentives of risk absorbers at the time as opposed to in advance (see \cite{klagesmundt2018basis} for a critique).

We refer to a second type as \emph{miner-absorbed} (e.g., \cite{Kowala}), which aims to stabilize the base asset of a blockchain by manipulating protocol incentives. These designs propose for the supply to be dynamically adjusted by manipulating mining rewards, mining difficulty, and the level and burning of transaction fees or interest charges. This means that miners take an implicit risk absorber position that is meant to absorb price risk, but without an obligation to continue mining/risk absorbing. In many ways, this parallels the Basis/Nubits design. Miners are rewarded with newly minted stablecoins when the supply needs to be increased and face slashed rewards and burned transaction fees if they choose to continue mining when the supply needs to be reduced.

\subsection{Risk Absorption and Issuance}

The stablecoin mechanism works when speculators are incentivized to absorb price risk. These risk absorbing positions have two primary forms. In \emph{equity risk absorption}, a secondary asset exists, and any holder of this asset implicitly absorbs risk from the stablecoin. For instance, the Steem market cap implicitly backs Steem Dollars; a Steem Dollars holder can redeem Steem Dollars for newly minted Steem, and all Steem holders bear this inflation cost. In \emph{agent risk absorption}, individual agents manage a vault containing primary value that absorbs stablecoin risk. In agent risk absorption, agents decide how much to participate with their asset whereas, in equity risk absorption, every holder of the secondary asset participates proportionately. In many cases, the risk absorber role is also combined with stablecoin issuance.

An issuance process determines the stablecoin supply. A lot of variation is possible in the process specifics, but there are two general types. In \emph{agent-based issuance}, the size of the stablecoin supply, or more specifically the leverage of the system (the size of the stablecoin supply relative to the collateral value), is decided by agents in the course of optimizing their positions. The deciding agents are typically the risk absorbers in the system. For instance, in Maker, vaults determine their stablecoin issuance in managing the leverage of their vaults. In NuBits, owners of `equity'-like shares collectively vote on issuance decisions to balance demand.

In \emph{algorithmic issuance}, a process to adjust leverage (relative supply) is codified in the stablecoin protocol. For instance, in Duo Network, leverage is determined algorithmically through `leverage resets', which balance the stablecoin supply relative to collateral value. In seigniorage shares, new issuance is awarded algorithmically to `equity' holders to balance demand.

A \emph{deleveraging process} is also part of issuance that can be invoked to reduce the stablecoin supply if a deleveraging factor is breached, or if stablecoin holders are allowed to redeem stablecoins for the collateral. For instance, in Maker, if the stablecoin issuance of a vault is too large relative to the collateral value, the collateral is liquidated to reduce leverage. In Duo Network, `leverage resets' may force the liquidation of some positions if a collateral factor is breached. In seigniorage shares, losses are born by `equity' holders to reduce the stablecoin supply in a demand shock. In Steem Dollars, if price is below target, stablecoin holders may redeem for newly minted Steem.

As introduced in \cite{klagesmundt2019} and \cite{klagesmundt2020}, non-custodial stablecoins based on leveraged lending markets face deleveraging risks, which can cause feedback spirals on primary value. Most existing non-custodial stablecoins fit this leveraged lending characterization. 
These deleveraging risks take two forms. 
The first is a feedback effect on the stablecoin market: collateral value may be consumed faster in liquidations due to drying of stablecoin liquidity. 
The cost of deleveraging in a crisis may be significantly higher than \$1 per stablecoin due to this effect, as predicted in \cite{klagesmundt2019} and validated in Maker during `Black Thursday' in March 2020. 
The second is a feedback effect directly on endogenous and implicit collaterals.
For endogenous collateral, liquidations can cause a liquidity and fire sale effect on the collateral asset market in addition to a feedback effect on reduced expectations.

A similar feedback occurs in implicit collateral and affects the risk absorbers' positions and stablecoin demand. For both types of implicit collateral, there is a ceiling on how much can be absorbed. For seigniorage shares, this is in demurrage of equity holders. For miner-absorbed, this is likely around 0 block reward, except possibly in staking systems in which stake can be slashed as demurrage. The result is feedback in the participation incentives and value of the risk absorbing position. For instance, for miners to be willing to continue mining without a mining reward, the expectations of future profit need to outweigh the costs. A continued participation decision will depend on whether the investment can be repurposed and potential returns from competing alternatives. After this ceiling, the remaining flexibility is only in burning of fees charged in stablecoin usage, which has a feedback effect on the attractiveness of holding the stablecoins.

This leads to two universal and fundamental questions:
\begin{question}[Incentive Security]
\label{q:incentive-security}
Is there mutually profitable continued participation across all required parties?
\end{question}

If not, then the mechanism cannot work as no one will participate. This question also includes incentives around attacks; in particular, if incentives lead to profitable attacks, then \emph{rational} agents will be less inclined to participate. After this is answered, we can then make sense of the follow-up question:
\begin{question}[Economic Stability]
\label{q:economic-stability}
Do the incentives actually lead to stable outcomes?
\end{question}


Note that particular feedback effects can be mitigated. However, the result is typically to shift the risk from one agent to another. In either case, the risk will affect participation incentives. For instance, in collateral liquidations, some stablecoin holders could be liquidated at par for the collateral asset as opposed to at a floating market price. This eliminates the feedback effect on the stablecoin market price, reducing deleveraging risk on risk absorbers. Instead, however, the stablecoin may be less attractive to stablecoin holders as they now take on more liquidation risk.

The type of stablecoin structure will also significantly affect incentives. When designs are more agent-based, agents have greater decision flexibility and are more likely to find a profitable participation level. In comparison, when designs are more algorithmic and/or with equity risk absorption, agents are more restricted and may be less likely to participate in the system relative to alternatives.\footnote{An interesting anecdote is the `miracle' of the W\"{o}rgl Experiment. In this experiment, currency demurrage is purported to stabilize the local economy in a depression by incentivizing current spending. However, as discussed in \cite{goodwin2013}, this ought to have an effect on participation incentives, leading to a lower equilibrium price of the demurrage currency relative to alternatives.}
Several past stablecoin events serve as case studies for deleveraging effects. These are described in Table~\ref{table:deleveraging_events} in the Appendix.

Stablecoins can also incorporate other insurance mechanisms to mitigate risk (e.g., ~\cite{opyn,yinsure,nexusmutual}).
The simplest is creating a fully collateralized put option market, from which individual stablecoin holders can purchase an option to swap from this stablecoin to another stablecoin/asset. 
Naturally, this insurance is only as valuable as the collateral behind it. 
Other insurance mechanisms add a layer to the protocol intended to globally buffer against shortfalls---e.g., in case the `dynamic' part of the CDO structure fails to cover all losses.
In some cases, these can be interpreted as a `mezzanine' tranche in the CDO-like structure, though this is not completely accurate as this `tranche' is often unsecured. 
In particular, many current stablecoins generate cash flows from fees that are securitized into governance tokens (e.g., MKR in Maker)).
To cover a shortfall situation, the value of future cash flows can be auctioned off by selling new governance tokens. However, the value of future cash flows can evaporate in death spiral situations. Alternatively, a portion of past fees can be diverted to serve as a buffer to cover shortfalls. There is in fact a spectrum between these options, in which securitized cash flows can be sold at arbitrary times to maintain an adequate buffer.\footnote{This can be interpreted similarly to corporate financing decisions around if/when to raise capital vs. internally finance.}

\paragraph{A design gap: buffers}
This largely unexplored spectrum of options represents a more general design gap: an under-appreciation of buffers in stablecoin design. 
\cite{klagesmundt2020} shows that leveraged lending-based stablecoins can be stable in regions in which the underlying collateral price process is a submartingale (i.e., the next period expected return is positive) and can break down outside of this.
While there is some concern about the reasonableness of a submartingale assumption, it may be more reasonable in a relaxed form, in which downward movements are transitory (or long-term expected return is positive). 
There is little that derivative design can do to help systems survive aside from transitory downside events. 
In this relaxed form, it is important that systems have adequate buffers so as to survive transitory events; we suggest that many concerns about the appropriateness of submartingale assumptions can be translated to concern about adequate buffer size.
In this way, we expect an optimized buffer design can extend regions of stability for stablecoins, whereas this is largely underexplored in current designs.\footnote{For instance, Maker has a `system surplus’ account that served as a buffer during Black Thursday. This was not in fact intended as a stability buffer and is typically used to accrue fees until they reach a size for returning to `equity’ holders. Instead, Maker’s intended buffer is an auction of MKR, arguably at the worst possible times, to cover shortfalls.}
Another form of such a buffer is proposed in \cite{klagesmundt2020}: vault insurance that can cushion the effects of deleveraging spirals.

We also suggest that well-designed buffers can expand design possibilities beyond leveraged lending-based stablecoins. For instance, stablecoin designs with different fundamentals based on money market fund and currency peg models where the peg is maintained by an internal buffer effect. One example of these ideas is discussed more in the context of \emph{composite} stablecoins in Section~\ref{sec:dis_composite} and in \cite{klagesmundt_grant2018,klagesmundt-ic3}.



\subsection{Governance, Mining, and Manipulation}
We now introduce design components that introduce manipulation potential in the system. In custodial systems, such manipulations are typically avoided by relying on societal institutions. In contrast, permissionless systems usually do not offer strong identities, which open up various anonymous attacks that cannot be prevented by institutions. The precise form of these components affect the size and scope of attack vectors, but don't substantially change their form; thus we focus our discussion on the functional forms that are important for economic models. We provide a list of historical manipulation events as case studies in Table~\ref{table:manipulation_events} in the Appendix.

\paragraph{Data Feeds}
Non-custodial stablecoins require asset price data in terms of the target peg (e.g., ETH/USD prices). This data is not natively accessible on-chain since fiat-cryptocurrency conversions can only take place on off-chain exchanges. As a result, the stablecoin relies on a mechanism to import this data into the blockchain virtual machine so that it is readable by the stablecoin smart contracts (also known as an `oracle'). As a result, the correctness of the imported data is not objectively verifiable on-chain, as opposed to native actions such as intra-blockchain transaction validity or inter-blockchain transaction validity~\cite{Zamyatin2019XCLAIM}. There are various methods, both centralized and decentralized, to construct such data feeds. We give a brief overview of these in the appendix. Though, from a functional standpoint, we can abstract from the technical details to focus on the economic structure that these data feeds add.

Data feeds introduce a new incentive problem: if importing data into the system has an extractable value X, then an attacker will spend up to X to manipulate that data. Centralized feeds can be manipulated by the counterparty, which introduces potentially perverse incentives for the counterparty as well as single points of failure. Decentralized methods typically collapse in the face of game-theoretic attacks. As a result, data feeds add an inherent manipulation potential into our general model. \emph{The important factors of this include who can manipulate the feed, how much the feed can be manipulated, and the cost involved in such manipulation.} Given this, a reasonable aim is to achieve data feed incentive compatibility to report honestly in the combined data feed-stablecoin system.

\paragraph{Governance}
Stablecoin governance is tasked with managing system parameters, such as interest rates, collateral factors, data feed curation, time delays, system upgrades, and emergency system settlement. In return, they typically receive some fee revenue from the system. Governors may take the form of governance token holders who vote on parameters, the founding company, a subsumed role of other agents in the system, or may be algorithmic.

If it is performed by agents, then these agents have power to manipulate the system through these parameters. For the system to be secure, governance must be disincentivized from fatally attacking the system. The potential for profitable attacks will feedback into the participation decisions of the other agents in the system. For instance, if governance is tokenized, then the token valuation/expectations, which could be slashed after an attack, and any other costs must be sufficiently higher than the proceeds of the attack. We discuss several attacks, involving manipulations of data feeds and parameters to extract collateral value, in the context of proposed models in the next section.

Governance is also inter-related with system stability. In this anonymous setting, governance can be expected to maximize expected profits as opposed to targeting stability for its own sake, as is typically assumed in central bank models. It is an open question to what extent various governance structures align incentives with the targeting of stability.

On the other hand, if governance is algorithmic, the stablecoin may be susceptible to gaming attacks from the other participants. These attacks can take a related form assuming the governance algorithm as given and construct similar end results: e.g., bribe the chosen data feeds in order to extract system value. Potential profitability of these attacks will feedback into participation incentives of the agents in the system.

\paragraph{Miners}
A non-custodial stablecoin is implemented in a base blockchain layer. 
This can either be ``on top'' of a blockchain in the form of smart contracts or directly into the core runtime. In either case, the base blockchain is maintained by a set of miners. In this paper, we subsume both miners (typically used in the context of PoW) and validators (typically used in PoS) under the term ``miner''. In maintaining the blockchain, miners decide transaction inclusion and ordering in the ledger--both in the next block mined and in the previous blocks, as a miner could always choose to re-mine an earlier block to change the transaction structure. Hence, they have full control over the history of the ledger.

The blockchain system \emph{intends} for miners to ensure desired properties of persistence and liveness of the ledger~\cite{Garay2015}.
In this context persistence states that a valid transaction included in the ledger is eventually considered final, i.e., all honest agents will report the transaction in the same position in the ledger. 
The liveness property requires that a transaction sent from an honest agent is eventually inserted into the ledger.
In return, miners are paid a rewards in the form of fees for including transactions into blocks and block rewards for extending the ledger with new blocks.
Since present and future rewards are typically paid out in the base asset, miners have an incentive to avoid attacks that jeopardize these rewards.

However, miners can also receive payoffs from other sources outside of the blockchain protocol. For instance, miners can capture arbitrage opportunities in the exchange of assets on the ledger or by placing bets and manipulating the outcomes in the course of mining, or receiving bribes to do so on behalf of others~\cite{McCorry2018Bribing}. This is broadly summarized as Miner Extractable Value (MEV)~\cite{daian2019}. A rational miner will decide profit-maximizing actions taking MEV into account, which may not always be honest mining supporting the blockchain. If MEV is valuable enough, miners will generally be incentivized to capture it through an attack.

MEV poses a few risks in the context of stablecoins. First, specialized attacks are possible that exploit stablecoin deleveraging events and liquidations \cite{klagesmundt2019}. This leads to MEV opportunities that can incentivize destabilizing attacks on the stablecoin.
Understanding security and incentive alignment in this context and game theoretic interaction of many stablecoin agents and miners remain open problems.
Second, miner attacks pose consensus risk to the blockchain layer (e.g., affecting persistence). An attack of this form could have an effect on the base asset of the blockchain, which may be a collateral asset in the stablecoin. This can have an effect on stablecoin stability even if the stablecoin itself is not the focus of the attack.
Third, in the case of stablecoins embedded in the base protocol, the stablecoin may directly manipulate miner reward incentives, as opposed to indirectly manipulating incentives via MEV. This presents a related open problem of whether such blockchains can function (e.g., whether liveness is achievable).


\paragraph{Miscellaneous risks}
We briefly mention two other risks. One is often called `smart contract risk'. Since stablecoin systems execute algorithmically without specific institutional oversight, they face the risk of bugs in their specification and implementation--e.g., transaction-ordering dependencies, overflows, and re-entrancy. These risks may be representable in similar ways to credit risk models by introducing some probability of `default', in this case a software bug, and some random recovery ratio. Formal verification methods are typically used to mitigate these risks. Another risk is contagion risk from other protocols. In real environments, these systems do not occur in isolation. For instance, cascading liquidations in ETH and BTC between multiple leverage platforms occurred on `Black Thursday' in March 2020. We suggest that cascading liquidations like this can be modeled using fire sale models of networks of common asset holdings (e.g., \cite{braverman2018}).

\section{Models and Measures of Non-Custodial Stablecoins}
\label{sec:questions}

Based on the novel risks in non-custodial stablecoins, existing financial models cannot be used `out-of-the-box'.
Here we introduce foundational models for non-custodial stablecoins which adequately capture these risks.
First, we draw inspiration from capital structure models, extending a basic model to capture additional aspects and formulate four formal examples of such problems.
Second, we consider forking models, moving from the single-shot nature of the capital structure models we present to games of multiple rounds. 
Third, we provide a brief review of models that focus on whether non-custodial incentive structures can lead to stable price dynamics.
Finally, we include an estimation of utility functions specifically for the Maker protocol.






\subsection{Capital Structure Models}

We draw inspiration from capital structure models (\cite{dybvig1991}, \cite{myers1984}) to understand incentives and attacks in stablecoins. The original formulation of these models describe incentives in an IPO offering between equity holders, bond holders, and managers. In the stablecoin adaptation, the model describes incentives between governors who hold governance tokens ($\sim$ equity), stablecoin holders ($\sim$ bond holders), and vaults/risk absorbers ($\sim$ managers). We relate vaults to managers as vaults decide the stablecoin supply.

We consider three assets: COL (collateral asset, e.g., ETH), GOV (governance token), and STBL (stablecoin). In Problems \ref{prob1}-\ref{prob2}, we consider vaults endowed with COL, governors endowed with GOV, and stablecoin holders who purchase STBL. In Problem~\ref{prob3}, we consider a different formulation in which agents choose portfolios of assets, including strategic holdings of GOV. We define the following model components
\begin{itemize}
	\item $N$ = dollar value of vault collateral (COL position)
	
	\item $R$ = random return rate on COL
	
	\item $F$ = total stablecoin issuance (debt face value)
	
	\item $b$ = return rate on a new opportunity; vault issues stablecoins (raises debt) to pursue this
	
	\item $\beta$ = collateral factor
	
	\item $\delta$ = interest rate paid by vault to issue STBL
	
	\item $u$ = vault's utility from an outside COL opportunity 
	
	\item $U(\cdot)$ = stablecoin holder's utility function
	
	\item $B$ = STBL market price at issuance
	
	\item $P_t$ = GOV market value at model time $t$ with terminal valuation parameter $\kappa$.
\end{itemize}

The model proceeds in three stages: (0) governance decides interest rate $\delta$ (i.e., the contract with the vault), (1) vault decides stablecoin issuance leveraged against a collateral position, and (2) the system is settled with an attack occurring if profitable. In a simplest formulation, the vault and governance are assumed to maximize expected value (risk neutral), and the stablecoin holder has risk averse utility $U$ with unlimited demand depth at this utility, which we later relax.

The three model stages lead to a sequence of GOV token prices $[P_0, P_1, P_2]$. In the simplest form, these represent discounted cash flows accruing to governance given the information at each time. Note that which $P_t$ appear in an optimization problem will depend on the precise problem setting we model. $P_0$ is the objective that governors optimize in period 0. $P_1$ gives the GOV valuation after vaults and stablecoin holders strategically participate in GOV ownership (e.g., in Problem~\ref{prob3}). $P_2$ gives the GOV valuation at the end of the model. Conditioned on no attack taking place, $P_2 = \delta F + \kappa$, where $\kappa$ is a terminal valuation parameter. If an attack occurs, then we assume participants abandon the system yielding $P_2 = 0$. The terminal valuation $\kappa$ represents the growth potential of the stablecoin: for instance, if $F$ becomes large in the future, then GOV cashflows $\delta F$ become large as well.

\subsubsection{Problem~\ref{prob1}: Capital structure with no attack}
Problem~\ref{prob1} introduces a simple setup with no attacks.
This resembles the classic capital structure problem (and can be solved similarly to \cite{dybvig1991}) with a particular form of contract between the equity and manager: now, vaults receive all profits from leverage with an interest fee paid to governance. The governance choice problem is to maximize the expected fee revenue subject to the vault's stablecoin issuance. The vault choice problem is to maximize expected returns from leverage minus fees subject to these constraints: (1) the collateral constraint, (2) the participation constraint, (3) stablecoin market price as the stablecoin holder's expected utility of holding one stablecoin. 

Notice that, for simplicity, there are several limitations to the model as formulated. In a more complete model, the vault may account for collateral liquidation costs (as in \cite{klagesmundt2020}) and last-resort insurance roles of GOV to make up for any collateral shortfalls (which can be accounted for by adding terms of $- [F(1+\delta) - N(1+R)]^+$ to the governance objective and modifying the stablecoin pricing constraint). Some stablecoins also include an interest rate paid to or by stablecoin holders. Finally, notice that both the setups with sequential choices by the vault and the governance as well as concurrent choices are realistic.

\begin{problem}
\footnotesize
	\begin{align*}
	\multispan2{\underline{\textbf{Governance choice}}}\\
	\max_{\delta\in[0,1)} \hspace{0.5cm}	& \EX \Big[ \delta F + \kappa  \Big] \\
	\text{s.t.} \hspace{0.5cm}	& F \text{ is vault choice}
	\end{align*}
	\begin{align*}
	\multispan2{\underline{\textbf{Vault choice}}}\\
	\max_{F\geq 0} \hspace{0.5cm}	& \EX [ NR + F(B b - \delta) ] \\
	\text{s.t.} \hspace{0.5cm}
	& F \leq \beta N \\
	& u \leq \EX [ NR + F(B b - \delta) ] \\
	& B = \EX\Big[ U\Big( \frac{1}{F}\min( F, N(1+R) - \delta F ) \Big) \Big]
	\end{align*}
	\caption{Capital structure with no attack vectors}\label{prob1}
\end{problem}


\subsubsection{Problem~\ref{prob2}: Capital structure with governance attack}

We consider a governance attack vector of the form described in \cite{zoltu2019} and \cite{gudgeon2020}. In such an attack, an agent with a $\zeta$ fraction of GOV tokens is able to steal $\gamma$ fraction of collateral in the system. As described in \cite{zoltu2019}, this could occur in the Maker system at the time with $\zeta=0.1$ and $\gamma=1$ (or possibly $\gamma > 1$ after accounting for simultaneous attack on other systems using the stablecoin) because governance is granted the power to arbitrarily alter the contracts.\footnote{Note that governance attacks like this can be mitigated by limiting the contract structure governance can alter and implementing long time delays between changes, but it is a realistic attack vector in currently deployed systems that build in broad contract upgrade capability. The structure of the formal problem can also be altered by tailoring emergency settlement triggers.}

This attack is profitable if the proceeds exceed the costs:
$$\gamma N(1+R) > \zeta(\delta F + \kappa) + \alpha,$$
where $\alpha$ incorporates an outside cost to attack and $\zeta(\delta F + \kappa)$ is the opportunity cost of attack (the value of $\zeta$ fraction of GOV tokens). Note that in traditional financial settings, we typically have $\alpha >> \gamma N$: $\alpha$ represents a high cost due to legal/reputational recourse. This simplifies the problem to Problem~\ref{prob1} as the attack is always unprofitable.

In the Problem~\ref{prob2} setting, the governors split into two groups: attack and non-attack groups. If we think of individual governors having individual $\alpha$ costs to attack, then the attack group will form from the $\zeta$ fraction with lowest $\alpha$. If we take $\zeta < 0.5$, then the non-attack group will decide interest rate $\delta$ while the attack group will decide $d\in\{0,1\}$ whether to attack. If $\zeta > 0.5$, then the attack group decides both $\delta$ and $d$. Problem~\ref{prob2} models the case of $\zeta < 0.5$: the governance choice problem represents the non-attack group decision over $\delta$, and the attack group decision is represented by the $\Ind_d$ constraint. Note that a simple reformulation of the governance objective would model the case of $\zeta > 0.5$.

The vault decision is expanded to include the amount of collateral $N$ locked in the stablecoin subject to an amount $\bar N$ available to the vault; the amount locked is subject to seizure by a governance attack. This compares to Problem~\ref{prob1}, in which all vault COL is locked since there is no attack vector (the previous $N$ is the new $\bar N$). For simplicity, the setup assumes that $\gamma$ is such that, under a successful attack, no collateral is recoverable by the vault after accounting for $F$; this could be relaxed with an extra term in the vault's objective. As an extension to Problem~\ref{prob2}, $\alpha$ could also incorporate a bribe decision from the vault to governance to change attack incentives.

\begin{problem}
\footnotesize
	\begin{align*}
	\multispan2{\underline{\textbf{Governance choice}}}\\
	\max_{\delta\in[0,1)} \hspace{0.5cm}	& \EX \Big[ (1-d)\Big(\delta F + \kappa \Big) \Big] \\
	\text{s.t.} \hspace{0.5cm}
	& d = \Ind_{(\gamma N(1+R) > \zeta (\delta F + \kappa) + \alpha)} \\
	& F \text{ is vault choice}
	\end{align*}
	\begin{align*}
	\multispan2{\underline{\textbf{Vault choice}}}\\
	\max_{N, F\geq 0} \hspace{0.5cm}	& \EX [ (\bar N - N)R + (1-d)NR + F(B b - \delta) - d N(1+R)  ] \\
	\text{s.t.} \hspace{0.5cm}
	& F \leq \beta N \\
	& \Ind_{(N>0)} u \leq \EX [ F(B b - \delta) - d\gamma N(1+R)  ] \\
	& B = \EX\Big[ U\Big( \frac{1}{F} \min\Big( F, (1-\gamma d)(N(1+R) - \delta F) \Big) \Big) \Big] \\
	& d = \Ind_{(\gamma N(1+R) > \zeta (\delta F + \kappa) + \alpha)} \\
	& 0 \leq N \leq \bar N
	\end{align*}
	\caption{Capital structure with governance attack vector}\label{prob2}
\end{problem}

In Problem~\ref{prob2}, incentive alignment against attack (security) will depend critically on $\kappa$ and $\alpha$ as it's unrealistic for $\delta F$ to be on the order of $N$ ($\sim 100\%$ interest rate). In a long-run growth equilibrium $\kappa$ will be related to the geometric sum $\frac{\delta F}{1-r}$ for some discount factor $r$. This allows us to understand the settings in which long-run incentive security will depend on a large $\alpha$ term, which equates to centralized recourse. In particular, combining the conditions for a non-attack decision with the collateral constraint, we need $\frac{\gamma r}{\zeta \delta} < \beta$ to have incentive security against attack with $\alpha=0$, which is very limiting for practical values of these quantities. Notice that, if incentive security is lacking or the opportunity is not profitable enough for the vault, an equilibrium can be no participation from the vault (in which case $\Ind_{(N>0)}=0$ in the utility threshold constraint).

We can interpret this as a `price of anarchy' concept. In this case, we may want to measure the ratio between the `best decentralized equilibrium' and the optimal `centralized' solution (e.g., when $\alpha >>0$ simplifies the setting to Problem~\ref{prob1}). A natural task of a protocol designer would be to optimize this cost.

\subsubsection{Problem~\ref{prob3}: Portfolio selection with collusion attack}

We now consider a collusion attack vector of the form described in \cite{klagesmundt2019vuln}. For instance, a group that controls a large share of GOV (e.g., $51\%$, though possibly lower) can manipulate price feeds and settle the system such that stablecoin holders or vaults have claim to greater share of collateral. If the group also holds the profitable position (e.g., stablecoins), then the attack can be profitable unless the GOV token holds adequate market value. These $51\%$-style attacks can't inherently be mitigated.\footnote{Common mitigations include governance delays and maximum governance changes, but these are only effective to a certain extent. As discussed in \cite{klagesmundt2019vuln}, once there is a profitable coalition, they can wait out any time delays--e.g., vaults are not able to exit if they can't buy back the stablecoins.}

We model these attacks in a more complex setting; a full formal setup is in Appendix Problem~\ref{prob3}. In this setting, vaults and stablecoin holders are endowed with a value and choose a portfolio of available assets, some of which entail participation in the stablecoin system and are subject to attack. They may strategically bid up the price of GOV to secure the system or acquire GOV and/or issue a bribe to try to trigger a instigate a profitable attack. A third agent is an outside GOV holder who may choose to collude with other agents. These agents make the following strategic decisions:
\begin{itemize}
	\item Vault decides portfolio $\mathbf x$ allocated between COL and GOV, level of participation in the stablecoin $N$ and $F$, and bribe factor $\gamma_v$ to the outside governors.
	
	\item Stablecoin holders decide portfolio $\mathbf y$ allocated between STBL, GOV, and COL and bribe factor $\gamma_s$ to the outside governors.
	
	\item Outside governors hold $\varepsilon$ fraction of GOV, decide interest rate $\delta$ and decide whether to collude with the vault ($d_v$), the stablecoin holder ($d_s$), or whether no attack occurs ($d_n$).
\end{itemize}
The offered bribes are a $\gamma_v$ and $\gamma_s$ fraction of attack profitability. An attack is profitable if $\zeta$ fraction of governance collude (e.g., a threshold to manipulate the price feed)--we can generally take $\zeta \geq 0.5$, but could be lower if collusion with miners is added in. The portfolios $\mathbf x, \mathbf y$ have components measured in dollar value and which sum to the total endowed values $\bar x, \bar y$.

The COL market is assumed to be perfectly liquid at the given price, and so portfolio decisions have no price effect on COL. We restrict the focus to modeling endogenous prices of GOV and STBL. The price of GOV is determined through the function $P(\mathbf x_G, \mathbf y_G, \delta, F)$; we assume this $=\EX[\delta F + \kappa]$ without vault or stablecoin holder participation in the GOV market. In the model, $P_2 = P_1$ conditional on no attack. If an attack occurs, then GOV price goes to zero. The STBL price is determined through the function $B(F,\mathbf y_S)$ in a way that balances supply and demand. Since the stablecoin holder has an endowed value in this problem, we no longer assume the STBL market demand has an unlimited depth at a given utility value, as done in the previous formulations. The behavior of this model will likely depend largely on the choice of functions $P,B$. A number of choices could be explored to consider different market structures.

\setcounter{problem}{2}

\begin{problem}
\footnotesize
	\begin{align*}
	\multispan2{\underline{\textbf{Outside governance choice}}}\\
	\max_{ \scriptscriptstyle \delta\in [0,1) , \scriptscriptstyle d_{\{n,v,s\}}\in\{0,1 \}}
	\hspace{0.5cm}	& \EX \Big[
	d_n \varepsilon (\delta F + P_1 ) 
	+ d_v \big( \gamma_v(F-\mathbf x_G) - \alpha\big)\\
& \hspace{1cm}	+ d_s \big( \gamma_s (N-\mathbf y_G) - \alpha\big)
	\Big] \\
	\text{s.t.} \hspace{0.5cm}
	& P_1 = P(\mathbf x_G, \mathbf y_G, \delta, F) \\
	& \Ind_{(\frac{\mathbf x_G}{P_1} \geq \zeta)} \leq d_v \leq  \Ind_{(\varepsilon + \frac{\mathbf x_G}{P_1} \geq \zeta)} \\
	& \Ind_{(\frac{\mathbf y_G}{P_1} \geq \zeta)} \leq d_s \leq  \Ind_{(\varepsilon + \frac{\mathbf y_G}{P_1} \geq \zeta)} \\
	& d_n = (1-d_v)(1-d_s) \text{ and } d_v = (1-d_n)(1-d_s) \\
	& \mathbf x, \mathbf y, N, F, \gamma_v, \gamma_s \text{ from vault and stablecoin holder choices}
	\end{align*}
	\begin{align*}
	\multispan2{\underline{\textbf{Vault choice}}}\\
	\max_{\mathbf x, N, F\geq 0, \gamma_v\in[0,1)} \hspace{0.5cm}	& \EX \Big[
	\mathbf x_C R + F(Bb - \delta) + d_n \frac{\mathbf x_G}{P_1} (\delta F + P_1) \\
	& \hspace{1cm} + d_v (1-\gamma_v)(F-\mathbf x_G) - d_s N
	\Big] \\
	\text{s.t.} \hspace{0.5cm}
	& \Ind^T \mathbf x = \bar x \\
	& 0 \leq N \leq \mathbf x_C \\
	& F \leq \beta N \\
	& \Ind_{(N>0)} u \leq \EX \Big[ F(Bb - \delta) + d_n \frac{\mathbf x_G}{P_1} (\delta F + P_1) \\ 
	& \hspace{1cm} + d_v (1-\gamma_v)(F-\mathbf x_G) - d_s N  \Big] \\
	& B = B(F,\mathbf y_S) \\
	& P_1 = P(\mathbf x_G, \mathbf y_G, \delta, F) \\
	& \delta, d, \mathbf y \text{ from outside governor and stablecoin holder choices}
	\end{align*}
	\begin{align*}
	\multispan2{\underline{\textbf{Stablecoin holder choice}}}\\
	\max_{\mathbf y, \gamma_s\in[0,1)} \hspace{0.5cm}	& \EX \Big[ U\Big(
	\mathbf y_C R
	+ d_n \Big( \min\left( \frac{\mathbf y_S}{B}, N(1+R) - \delta F \right) + \frac{\mathbf y_G}{P_1} (\delta F + P_1)\Big) \\
	& \hspace{1cm} + d_s (1-\gamma_s)(N - \mathbf y_G)
	\Big)\Big] \\
	\text{s.t.} \hspace{0.5cm}
	& \Ind^T \mathbf y = \bar y \\
	& B = B(F,\mathbf y_S) \\
	& P_1 = P(\mathbf x_G, \mathbf y_G, \delta, F) \\
	& \delta, d, \mathbf x, N, F \text{ from outside governor and vault choices}
	\end{align*}
	\caption{Portfolio selection with collusion attack vector}\label{prob3}
\end{problem}

Compared to Problem~\ref{prob2}, the vault now decides the amount of COL to hold ($\mathbf x_C$), equivalent to previous $\bar N$) and, of that amount, the amount to lock as collateral in the stablecoin ($N$). Similarly, $\mathbf x_G, \mathbf y_G$ represents the amount of GOV in the vault and stablecoin holder portfolios respectively. We now have three attack decision variables ($d_n, d_v, d_s$), precisely one of which will take the value 1. The logic for this is encoded in the 2nd-4th constraints of the outside governance choice problem.

\subsubsection{Problem~\ref{prob4}: Miner-absorbed mechanism}
The miner-absorbed system is a variation of the presented problems as it explicitly models miners as the core participants.
The miner-absorbed stablecoin includes two agents:
\textit{Miners} taking the role of risk absorbers, governance and miners as well as \textit{stablecoin holders}.
Further, the system includes an algorithmic \textit{issuance} role (i.e., part of the base blockchain consensus protocol).
The primary value in a miner-absorbed mechanism is implicit collateral.
In this problem setting, we assume that miners are risk-neutral, economically rational agents\footnote{Non-risk neutral miners could also be observed and are covered for a non-stable currency in~\cite{Chen2019}}.
Further, we assume that the base blockchain includes a single currency STBL (i.e. the GOV and COL tokens are not present) and that it includes a correct and up-to-date price oracle.


We define Problem~\ref{prob4} as follows:
Should a miner generate a new block given an expectation of the rewards $r$ being paid, the return rate on the rewards $b$ at the market price of STBL $B$ considering the cost for mining $c$ as well as a long-term confidence in the system expressed as $P_1$?
In $c$ we subsume all variable and fixed costs for generating a block.
The miner's decision is expressed by $d$ such that $d=1$ encodes generating a block and $d=0$ the opposite.

The stablecoin holder decides to participate in the miner-absorbed systems based on the expected stability of the system expressed by the utility function $U$.
The stablecoin holder has a portfolio of assets $\mathbf{y}$.
The portfolio consists of two asset: STBL denoted as $\mathbf y_S$ and a second exogenous stablecoin denoted as $\mathbf y_A$.
For example, this could be a miner-absorbed system like Kowala and USDC as exogenous system.
The stablecoin holder re-balances the weight of the portfolio from one block (denoted by $\mathbf{y_0}$ ) to the next block (denoted by $\mathbf{y_1}$).
The decision is based on the price of STBL expressed by $B$ and the price of the exogenous stablecoin denoted as $B_A$.
Additionally, there is a cost $\delta$ to acquire STBL.
The stablecoin holders portfolio re-balancing has an impact on the price $B$ expressed by the abstract function $B(r,\mathbf{y_1},d,P_1)$.
If the stablecoin holder sells significant amounts of his STBL holdings, this should have a severe implications for the price.
Last, we define the abstract function $P(\mathbf{y_S}, d)$ that determines the confidence in the system of the stablecoin holder.
For example, the stablecoin holder could short-term sell STBL without affecting the long-term confidence in the system.
This is similar to a stablecoin holder using STBL to e.g., pay bills but planning to keep using the system in the long-run.

Miner rewards $r$ are adjusted by the issuance algorithm.
The issuance algorithm is left abstract.
However, the objective of the issuance algorithm is to minimize the change in price $B$.
We note that in a PoW system the reward is constrained such that $r \leq 0$ since the issuance algorithm can in the worst-case pay zero rewards but not ``take-away" existing value.
In a PoS system this can be achieved by slashing PoS miners as well as in seigniorage share systems were miners additionally hold a risky asset such as COL~\cite{Sams2015}.
The issuance algorithm takes as inputs the price function, but has to assume that $d=1$.
The miner-absorbed problem adopts previous components and adds new ones as follows:

\begin{itemize}
    \item $c$ = cost for mining a block
    \item $\delta$ = cost to obtain a stablecoin
    \item $u$ = stablecoin holder's utility for an outside STBL opportunity
    \item $r$ = reward paid in the next block
\end{itemize}

\setcounter{problem}{3}
\begin{problem}
\footnotesize
	\begin{align*}
	\multispan2{\underline{\textbf{Miner (governance) choice}}}\\
	\max_{d \in \{0,1\}} \hspace{0.5cm}	& \EX \Big[ d(Bb r - c) + P_1\Big] \\
	\text{s.t.} \hspace{0.5cm} & d u \leq \EX[Bb r - c] \\
	& r \text{ is algorithmic issuance}
	\end{align*}
	\begin{align*}
	\multispan2{\underline{\textbf{Stablecoin holder choice}}}\\
	\max_{\mathbf {y_1}} \hspace{0.5cm}
	    & \EX\big[U(
	        \mathbf{y_1}_S B + \mathbf{y_0}_A *B_A + (\mathbf{y_0}_S-\mathbf{y_1}_S)B(1- \delta)
	        )\big] \\
	\text{s.t.} \hspace{0.5cm} & B = B(r,\mathbf {y_1}, d, P_1) \\
	& P_1 = P(\mathbf{y}_S, d) 
	\end{align*}
	\begin{align*}
    \multispan2{\underline{\textbf{Issuance algorithm}}} \\
    \min_{r \geq 0} \hspace{0.5cm} & |B(r, \mathbf {y_1}, 1, P_1) - 1|
    \end{align*}
	\caption{Miner choice with no attack vectors}\label{prob4}
\end{problem}


Given the problem~\ref{prob4}, $r$ depends on the the expectation the stablecoin holder has towards the price of STBL $B$ and the subsequent re-balanacing of the portfolio $\mathbf{y}$.
If the stablecoin holder expects the price stability, he will either increase his holdings of STBL (considering the cost of obtaining expressed by $\delta$) or keep his current holdings.
On the other hand, price instability will lead to a reallocation of portfolio weights towards the exogenous stablecoin\footnote{We note that we could extend this model with a preference for either STBL or the exogenous stablecoin.
For example, if the stablecoin holder prefers a non-custodial STBL and his only alternative would be a custodial exogenous stablecoin, we could increase the preference of STBL.}.
We discuss the changes in portfolio allocation as these lead to more severe impacts on $r$.

\textit{Case 1: Increased demand for STBL $\mathbf{y_{0_S}} < \mathbf{y_{1_S}}$.} 
To keep the price stable (i.e. $\min |B() - 1|$), the issuance algorithm sets $r>0$.
In turn, this increases the total supply $F$.
Assuming that $Bbr > c$, miners should choose to mine a block such that $d=1$.
Notably, the issuance algorithm can increase $r$ to meet any demand by simply increasing mining rewards.
However, there is can still be a problem here:
$r$ is directly paid to miners.
If miners are not spending STBL such that it is reallocated to stablecoin holders, even issuing $r$ can lead to a price increase.
Conversely, if $r$ is set too high and miners sell STBL directly, the price of STBL can decrease.
Hence, finding a price-stabilizing issuance algorithm is non-trivial given that the portfolio allocation and miner decisions cannot be known a priori.

\textit{Case 2: Decreased demand for STBL $\mathbf{y_{0_S}} > \mathbf{y_{1_S}}$.}
In this case, stablecoin holders are selling STBL in favor of an exogenous stablecoin.
The issuance algorithm reduces $r$ in return to limit the increase of $F$ or do not increase $F$ at all.
However, the problem of paying low rewards introduces two distinct problems.
First, it is possible that even in the case of $r=0$, $B$ will still decrease if there is too much supply in the market.
A short-term price increase might still be counter-acted if stablecoin holders and miners have long-term confidence in the system expressed by $P_1$.
However, second, without block rewards, the expected utility for miners is can be negative since they cost for mining a block $c$ is only compensated with the long-term confidence $P_1$.
If miners only consider the next block (without $P_1$), the liveness of the ledger is sacrificed due to the ``Gap Game"~\cite{Carlsten2016, Tsabary2018}.
Even worse, miners could fork the chain with the most valuable transactions from the previous blocks to continue to earn rewards.
If the liveness of the miner-absorbed system is not present, it will likely also affect the long-term confidence in the system for stablecoin holders and miners.

Moreover, if the miner can easily switch between different chains, they would likely abandon the current stablecoin chain for one that pays high rewards.
One can motivate the miner to stay if the cost for switching is high, e.g., if a miner does not produce blocks in a given time they are slashed as in PoS systems.
However, hard-to-leave also means hard-to-join: a miner needs to be ensured that his rewards will be positive in expectation.
By adding up-front requirements like specialized hardware or acquiring certain currency, the rewards in expectation are minimized by the cost of acquisition as well as opportunity cost for maintaining the hardware/stake of coins.

\subsubsection{Further variations}

\paragraph{Endogenous collateral}
We now need to account for the endogenous COL price: the actions of the stablecoin agents will have a direct price effect on COL if the primary use of COL is within the stablecoin system. One way is to define the COL price return as a function of the decision variables and update the vault and stablecoin holder objectives with this price formulation. In this way, a driving random variable (like $R$ in the exogenous formulation) describing outside faith in the system would be an input to the price function in addition to agent decisions. As with the functions $B,P$ in Problems~\ref{prob1}-\ref{prob2}, the precise formulation of this price function will play an important role in the problem, but we can explore a number of different market structures.
In addition, the governance and vault roles may be merged into the same position if GOV = COL. Governance can also be an outside party without an explicit token--e.g., addresses controlled by the founding company.

\paragraph{Algorithmic issuance}
When stablecoin issuance is automated by the protocol, the vault is no longer a player. Instead, the issuance process becomes a constraint for the remaining players, as in Problem~\ref{prob4}. The issuance process will directly affect the value of GOV, in which case, it may be worth considering a participation decision in owning GOV (e.g., in a portfolio selection problem). If all COL is implicitly backing the stablecoin, an insurance role will factor into a general COL holder's decision to hold COL, and thus into the pricing of COL. If GOV = COL, then this all comes down to the pricing of GOV.
In the case that a specific portfolio of COL (and/or other assets) is backing STBL, and not all COL, then a money market model may be useful. Models such as \cite{parlatore2016} could be adapted to consider portfolio and last-resort insurance role of GOV ($\sim$ sponsor support) in a stablecoin setting with added attack vectors.

\paragraph{MEV: Miners as additional governance}
Some single period MEV attacks can be modeled within the capital structure framework by including miners as a second governance-type agent, who decides transaction inclusion and ordering. For instance, miners could earn potential profits from front-running STBL issuance decisions or from bribes to limit the actions of other agents. For richer MEV attacks, we describe the adaptation of blockchain forking models in the next section.

\subsection{Forking Models}

The capital structure models consider a single time-step: depending on the expectations of agents, they will choose to execute certain actions in the next round.
In this section, we extend the models to explore how multiple rounds of agent decisions can affect stability and security of stablecoin systems.
Specifically, we need to consider feedback mechanisms between different agents interacting over multiple rounds.
In such a setting, agents adjust their \emph{future} actions based on their beliefs of the other agents' actions and the output of the integrated algorithms (e.g., issuance or/and governance).
Moreover, we consider that permissionless ledgers used in non-custodial designs (e.g. Maker) lack finality.
Miners are able to re-order transactions and re-write history within certain depths of the ledger~\cite{Garay2015}.
This allows agents to adjust \emph{past} actions as well\footnote{While only miners can directly re-order and decide on the inclusion of transactions, other agents can employ bribing strategies to effectively achieve similar outcomes~\cite{McCorry2018Bribing}.}.
The resulting forking models are highly complex especially when considering a combination of a complex non-custodial system like Maker with a base blockchain like Ethereum.

Below, we consider a simpler formulation with specific couplings between otherwise separate models of a base blockchain and an application layer. An output of one layer would serve as exogenous input to the other layer and vice versa. For instance, the size of MEV determined in application layer participation feeds back into incentives for forking attacks in the base layer, which feeds back into the probabilities of attack in application layer incentives. In this way, a complex forking model could be simplified into simpler problems that can be solved iteratively to find an equilibrium.
This section is kept informal such that we describe the extensions required but do not include formal problems.


\paragraph{Base blockchain}
As explored in the blockchain folk theorem~\cite{Biais2019}, miners have an incentive to coordinate on the longest chain to increase their success of finding the next block.
However, if a miner is already invested in a fork, the miner decides based on his vested interest (e.g., accumulated work or committed stake) whether to switch to a different chain.
We need to take these two competing incentives into consideration when arguing about MEV, which serves as an implicit bribe for miners toward specific chains.
A forking model can explore the success probability of bribing miners based on their prior incentives.
Instead of modelling all miners with the same incentives, a forking model considers that miners already mining on a fork will have a higher incentive to take the bribe as they are invested in a fork.
Additionally, the setup in~\cite{Biais2019} can be extended by a network game as a stochastic dynamic system~\cite{dynamic-system} or a global game~\cite{globalgames} with noisy observations (e.g., network delay, reward expectations).
Moreover, we can incorporate various assumption of risk-appetite of miners~\cite{Chen2019}, selfish mining~\cite{selfish}, and the impact of block rewards in comparison to transaction fees~\cite{Carlsten2016,Tsabary2018}.  

\paragraph{Application layer}

A stablecoin that is built as an application on top of the base blockchain results in two directions of attack effects. In one direction, the application layer creates MEV that affects incentives on the base layer. For example, an agent wishing to prevent a liquidation transaction in Maker could offer a payment in another token to miners on Ethereum. Additionally, miners themselves are able to profit from their ability to determine the history of the ledger by e.g., execution of arbitrage opportunities, ``time-bandit attacks'', or oracle manipulation. Prior work on MEV in decentralized exchanges (DEXs)~\cite{daian2019} and data feed issues~\cite{zhang2016,chainlink} describe some effects of this direction.
The other direction affects participation in the application layer. A forking model could model the success probability of an exogenous bribe within the base blockchain. If successful, an attack would capture value locked in the stablecoin.
The possibility of such an attack (now or in the future) will have an effect on participation incentives in the stablecoin, similar to the description in the capital structure models. Stablecoin participation decisions in turn determine the size of MEV opportunities, which served as bribe inputs to the base layer model. Incentives created in the stablecoin system can therefore impact the security of the base blockchain system and vice versa.

\subsection{Price Dynamic Models}
We provide a brief review on models that explore the higher-level problem of whether non-custodial stablecoin incentive structures can lead to stable price dynamics. A challenge here is in modeling the feedback effects of agent decisions, as discussed in the previous section. To illustrate, in the most closely related traditional financial models, an assumed stable asset is borrowed against collateral, whereas in the non-custodial stablecoin setting, the `stable' asset that is borrowed has an endogenous price and/or participation level. The decisions of the other agents will affect this endogenous price and participation level of the stablecoin holder.

\cite{klagesmundt2020} and \cite{klagesmundt2019} construct stochastic models involving endogenous stablecoin price in exogenous collateral systems, taking into account deleveraging and liquidation actions given imperfectly elastic stablecoin demand. In this context, they model vault issuance incentives considering that issuance involves taking a leveraged bet on the collateral asset. They illustrate potential deleveraging feedback effects on stablecoin markets that lead to stablecoin price appreciation and characterize stable and unstable regions for stablecoins. As a result, vaults may have to pay above face value to deleverage in a crisis. This is validated by observed behavior of Dai on `Black Thursday', and was actually predicted a year before in \cite{klagesmundt2019}.

There are several open follow-up questions. For instance, evaluating the effect deleveraging events have on stablecoin holder participation incentives (particularly for different designs and relative to alternatives available to stablecoin holders), exploring strategic interaction of many vaults, destabilizing effects of attacks such as in the previously mentioned forking models, and extending to endogenous collateral models.

A few other papers are applicable to stability of stablecoins. \cite{gudgeon2020} and \cite{kao2020} model cryptocurrency-collateralized lending platforms. These do not incorporate feedback effects on the stable asset market, but do incorporate feedback effects on collateral asset liquidity.\footnote{These are similar to models for traditional collateral and debt security markets and repurchase agreements.} A simpler stablecoin problem involving no feedback effects is modeled in \cite{brown2019}. Option pricing theory is applied in \cite{cao2018} to value tranches in a proposed stablecoin using PDE methods, also under no feedback effects. Some stablecoins have also performed stability analyses (e.g., \cite{celo2019}, \cite{terra2019}), though these are typically limited in scope and include generous assumptions.

\subsection{Agents, preferences and attitudes to risk}

Agents' preferences, and in turn their behavior, are a central object in stablecoin design.
In Appendix~\ref{appendix:agents-preferences}, we first describe an framework which can be used to model preferences, and then outline two methods which can be used to estimate agents' risk attitudes.
The attainment of a clear understanding of agents' risk attitudes would serve to improve protocol design and parameter selection.

\section{From Stablecoins to DeFi}
\label{sec:discussion}

In this section we discuss a likely implication of our capital structure models.
Further, we outline how the modelling framework presented herein is applicable to other cryptoeconomic systems including composite assets, cross-chain protocols, synthetic assets, collateralized lending protocols, and DEXs.


\subsection{Sustainability of Incentives}
\label{sec:incentive-sustain}
As discussed in the context of our capital structure models, to maintain incentive security long-term, the value of a governance token may need to be disjoint from system growth. 
In particular, system growth rates (in supply, capital locked) are unlikely to be high in a long-term `steady state’ (and may be zero). 
However, the value of the governance token, if derived from discounted future fees, may only provide incentive security when the expected growth rates are high—in essence, when borrowing from the future is possible.
A long-term equilibrium without large future growth expectations may not be possible with governance token value derived from fees alone as they may be small with respect to value locked. 
Instead, other parties to the system may need to hold governance tokens to bid up governance token market value.
This will feedback into participation incentives of these other parties; there is no guarantee that equilibrium participation exists in this context either. 
To illustrate, stablecoin holders may need to hold significant positions in a risky governance asset in order to secure their stable positions, which may defeat their purpose in holding the stablecoin.
This leads us to a frustrating impossibility conjecture about many current systems in the context of our models:

\begin{conjecture}
In fully decentralized stablecoins ($\alpha=0$) with (i) multiple classes of interested parties (e.g., risk absorbers vs. stablecoin holders) and (ii) a high degree of flexibility in governance design, no equilibrium exists with long-term participation under realistic parameter values.
\end{conjecture}


An analogy helps to illustrate impossibility of some designs: if incentive security requires a bank's equity market value to be worth multiples of total deposits, then no depositors will participate. The bank's \emph{long-term} P/E ratio would need to be in the 100s or 1000s. The conjecture reinforces the importance of studying mutual incentives in choosing the right stablecoin design. Note that the oracle incentive compatibility problem also closely resembles the stablecoin governance incentive problem. Solving these problems in a fully decentralized way remains an open problem.

Current solutions implemented by stablecoins essentially centralize governance. This solution relies on a form of institutional liability and translates into a high $\alpha$ value (e.g., in Problem~\ref{prob2}). This is not necessarily a problem; many traditional financial systems operate in this way. This is why banks do \emph{not} need to be worth multiples of total deposits. However, we should openly recognize that this trust line exists and may be vital.



\subsection{Composite Stablecoins}\label{sec:dis_composite}
So far we have focused on \emph{primary} stablecoin mechanisms. 
Another class of \emph{composite} stablecoins involves baskets of primary stablecoins to try to further absorb risk.
The simplest is an \emph{ETF stablecoin}, which works using the ETF arbitrage mechanism to create/redeem the composite stablecoin against the basket.

A \emph{DEX stablecoin} aims to spread risk over the basket while providing an exchange service between the constituents, and so the basket weights change with exchange demand. 
DEX stablecoins take on the risk of liquidity provision to these exchanges. 
For constant function market maker (CFMM)-based exchanges, this risk is described in \cite{angeris2019,angeris2020improved}. 
Other DEX stablecoin designs propose limited 1-to-1 stablecoin swaps.
Existing DEX stablecoins bear the risk that the value of the basket may devolve into the value of the least valuable constituent(s) (e.g., if an underlying stablecoin fails).

A \emph{CDO composite stablecoin} segregates stablecoin risk into tranches.\footnote{Note the difference from the CDO analogy used to describe primary stablecoins.} 
For instance, the basket may have $n$ stablecoins and $n$ tranches. 
At settlement, the senior tranche holder gets first choice of which stablecoin to redeem for while holders of the most junior tranche picks last. 
Thus, junior tranche holders bear the risks of first stablecoin failures and are compensated with interest payments. 
This structure introduces a similar participation problem: enough agents need to be willing to take the different positions given the equilibrium level of interest payments.

A rainy day fund \emph{RDF stablecoin}, as introduced in \cite{klagesmundt_grant2018} and \cite{klagesmundt-ic3}, holds a basket of assets that accrues value to a safety buffer over time through arbitrage, fees, and other collateral uses. 
The collateral basket aims to target 1 USD, whereas the accrued buffer aims to smooth any asset failures/deviations over time.

Other composite stablecoins may also be possible. 
The stability of all composite stablecoins relies on primary stablecoin failures not being highly correlated.
Table~\ref{table:composite_category} summarizes categories for composite stablecoins, applicable models, and projects.

\subsection{Cross-chain and Synthetic Assets}
The foundations in this paper can also apply more broadly to synthetic and cross-chain assets.
In Appendix~\ref{sec:appendix-crosschain} we explain the relevant differences between these asset types in the present setting, and set out how our foundations apply.

\subsection{Lending Protocols and DEXs}

\paragraph{Lending protocols.}
Collateralized lending protocols share a similar structure to non-custodial stablecoins. 
Our models are easily adapted to describe such protocols. 
Lending protocols are simpler than non-custodial stablecoins in that borrowed assets are exogenous, rather than endogenously created by the protocol.
This makes system time delays more effective protective measures.
In the non-custodial stablecoin setting, a vault is not able to deleverage and exit unless they can repurchase stablecoins. 
Therefore in the event of a governance attack, a system time delay built into the protocol would likely be ineffective as a (profitable) coalition between stablecoin holders could simply wait out the delay, preventing many vaults from exiting.
In contrast, in the collateralized lending setting, an important security implication of the exogeneity of the borrowed assets is that it can allow protocol participants to leave a protocol before a governance attack is fully realized.
The typical borrowed asset either has a much larger market or is a custodial stablecoin, in which case the vault can always create new stablecoins at par through the issuer to deleverage. 
A system time delay could therefore protect participants by allowing them to exit before many impending governance attacks could be realized.\footnote{A likely exception is price feed attacks.}


\paragraph{DEXs.}
Some DEXs directly or indirectly have governance layers. 
When on the same native blockchain as the deposited assets, similarly to collateralized lending protocols, a DEX may also permit participants to exit before a governance attack is fully realized. 
However, where DEXs operate their own blockchain and control its governance (e.g., Rune), the ability for participants to exit in an attack can be fundamentally restricted. 
In this latter case, incentive security is an important question, and mutual participation of governance and other participants can be modeled as in our capital structure models.

For DEXs, fees are proportional to exchange volume while the potential payout of governance attacks is proportional to liquidity provider deposits. 
Therefore a key ratio of interest to protocol designers is volume relative to deposits. 
For a DEX, annualized volume can be as high as $\sim 100\times$ deposits (e.g. Uniswap). 
In comparison, for a collateralized stablecoin accruing fees on borrowed assets, such fees can be as low as $\sim 1/4$ of deposits. 
This $\sim 400\times$ factor makes the feasible region for incentive security against governance attacks potentially larger in DEXs than stablecoins.
This leads us to the following conjecture in the context of our models:

\begin{conjecture}
Considering fully decentralized systems ($\alpha=0$) with (i) multiple classes of interested parties and (ii) a high degree of flexibility in governance design, DEXs have a wider range of feasible long-term participation equilibria than stablecoins under realistic parameter values.
\end{conjecture}

An interpretation is that it may be fundamentally easier to economically secure DEXs against governance attacks than stablecoins. The conjecture also suggests ways in which broad stablecoin governance powers could be better aligned: by taxing transactions/economic activity ($\sim$ DEX volume) as opposed to assets under management. Of course, such a tax would make these stablecoins altogether less desirable to users with a cost for flexible governance.


\section{Concluding Remarks}
\label{sec:conclusion}

We have introduced a foundational framework for relating economic mechanics of all stablecoins and formulated three classes of models for non-custodial stablecoins, for which traditional financial models are sparse. These models evaluate measures of economic stability and incentive-based security considering mutual participation incentives of agents necessary for a mechanism to function. These models consider attack vectors including governance, data feeds, miners, and deleveraging market feedback effects.


\begin{acks}
We thank Andrew Miller and the anonymous reviewers for their feedback and suggestions.
This project received funding from a Bloomberg Fellowship, NSF CAREER award \#1653354, EPSRC Standard Research Studentship (DTP) (EP/R513052/1) and the BinanceX Fellowship programme. 
\end{acks}

\bibliographystyle{ACM-Reference-Format}

\appendix
\section{Appendix}
\label{sec:appendix}

\subsection{Tables}

\begin{table}[H]
    \begin{tabularx}{\columnwidth}{ccX}
        \toprule
    	\textbf{Category}	&   \textbf{Stability Models}    &	\textbf{Stablecoins} \\
    	\midrule
    	Reserve Fund	&   ETF &	TUSD, USDC, Libra v2 \\
    	Bank Fund		&   ETF, bank run  &	Tether \footnotemark[1] \\
    	MMF   &   ETF, MMF &   Libra v1 \\
    	CBDC	&   Currency    &	Chinese DC/EP \\ \bottomrule
    \end{tabularx}
    \caption{Custodial stablecoins and applicable models.
    NB as of 2019, Tether held 74\% reserves in USD/equivalents but claimed to be fully collateralized taking into account the value of loans to partner Bitfinex \cite{tether2,tether3}.}
    \label{table:custodial_category}
\end{table}

\begin{table*}
    \centering
    \includegraphics[width=\textwidth]{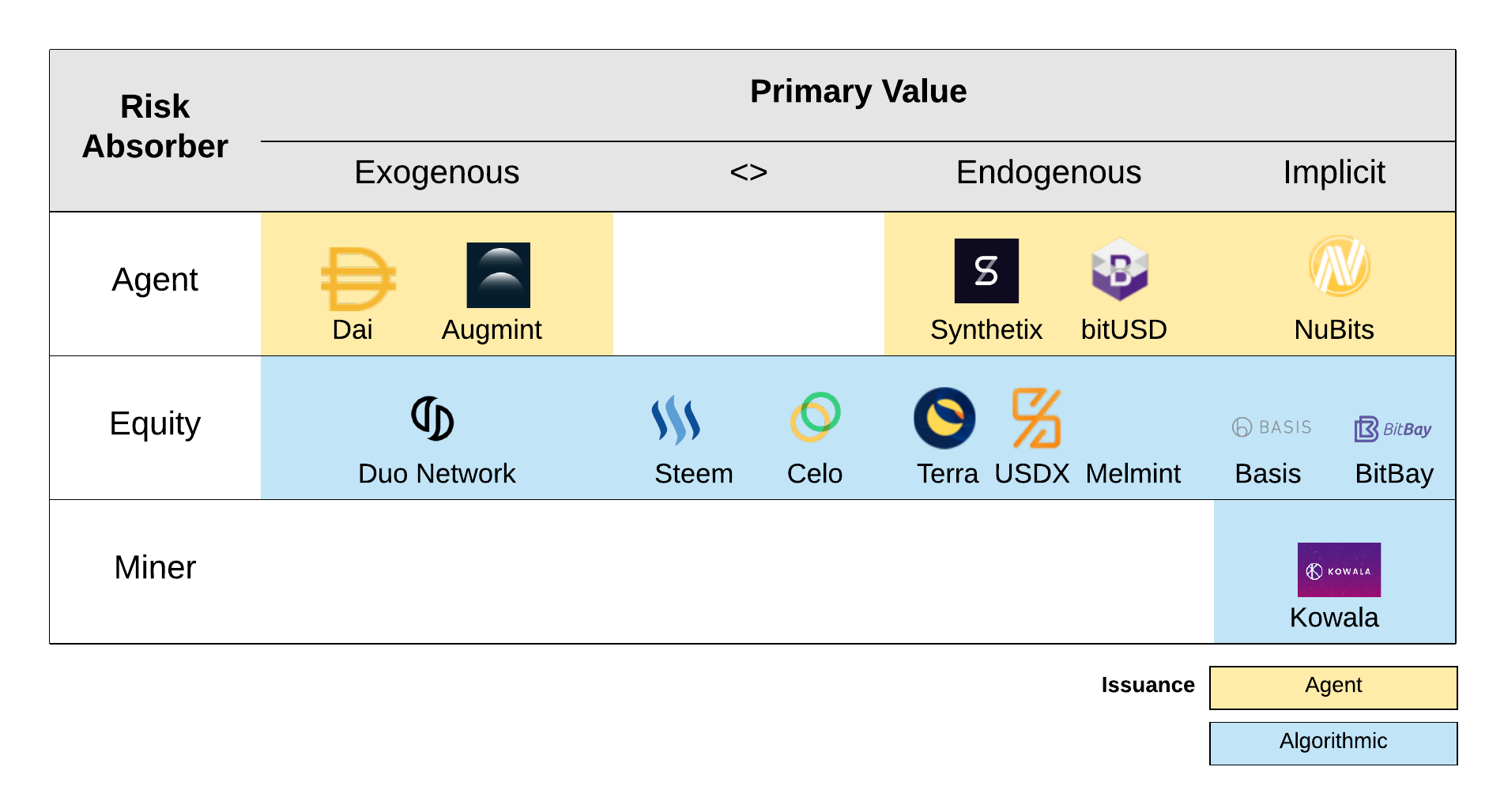}
    \caption{Non-custodial stablecoins as related by several components (excluding governance and data feeds).}
    \label{table:non-custodial_breakdown}
\end{table*}

\begin{table}[H]
    \begin{tabularx}{\columnwidth}{clX}
        \toprule
    	\textbf{Category}	&   \textbf{Relevant Models}    &	\textbf{Projects} \\
    	\midrule
    	ETF	&   ETF    &	Reserve \\
    	DEX &   Liquidity provider &   PieDAO, mStable, yCRV, CementDAO, Neutral \\
    	CDO &   CDO &   Introduced in \cite{buterin2018cdo} \\
    	RDF &       &   Introduced in \cite{klagesmundt_grant2018,klagesmundt-ic3} \\
    	\bottomrule
    \end{tabularx}
    \caption{Composite stablecoins summary.}
    \label{table:composite_category}
\end{table}

\begin{table*}
    \centering
    \includegraphics[width=\textwidth]{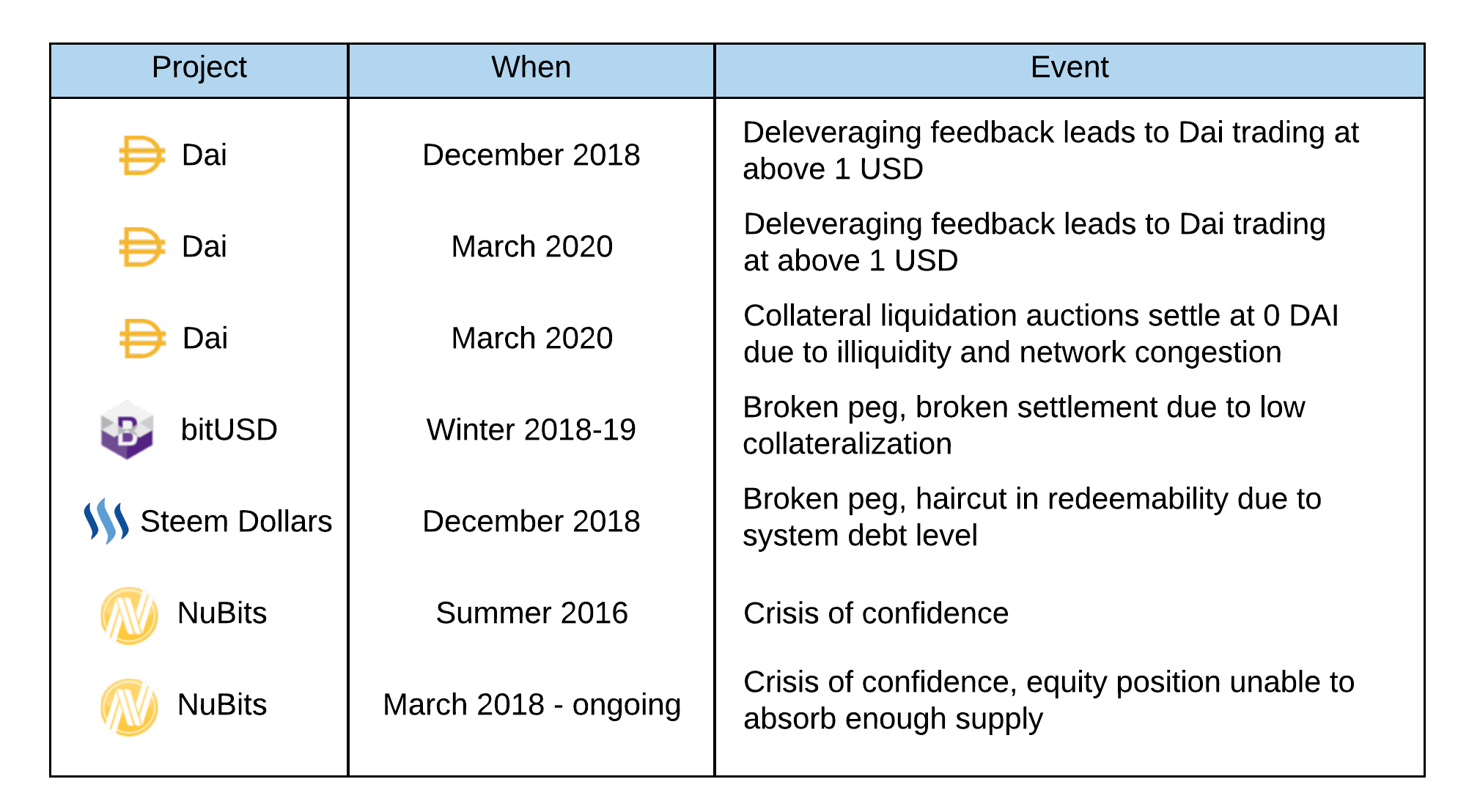}
    \caption{Notable non-custodial stablecoin deleveraging events.}
    \label{table:deleveraging_events}
\end{table*}

\begin{table*}
	\begin{tabularx}{\textwidth}{clX}
	    \toprule
		\textbf{Stablecoin}	&   \textbf{Time Period}    &	\textbf{Event} \\ \midrule
		Tether	&   Oct. 2018    &	Partner Bitfinex suspends fiat convertibility $\implies$ Tether crisis~\cite{tether} \\
		\bottomrule
	\end{tabularx}
	\caption{Custodial stablecoin depegging events.}
	\label{table:custodial_events}
\end{table*}

\begin{table*}
    \centering
    \includegraphics[width=\textwidth]{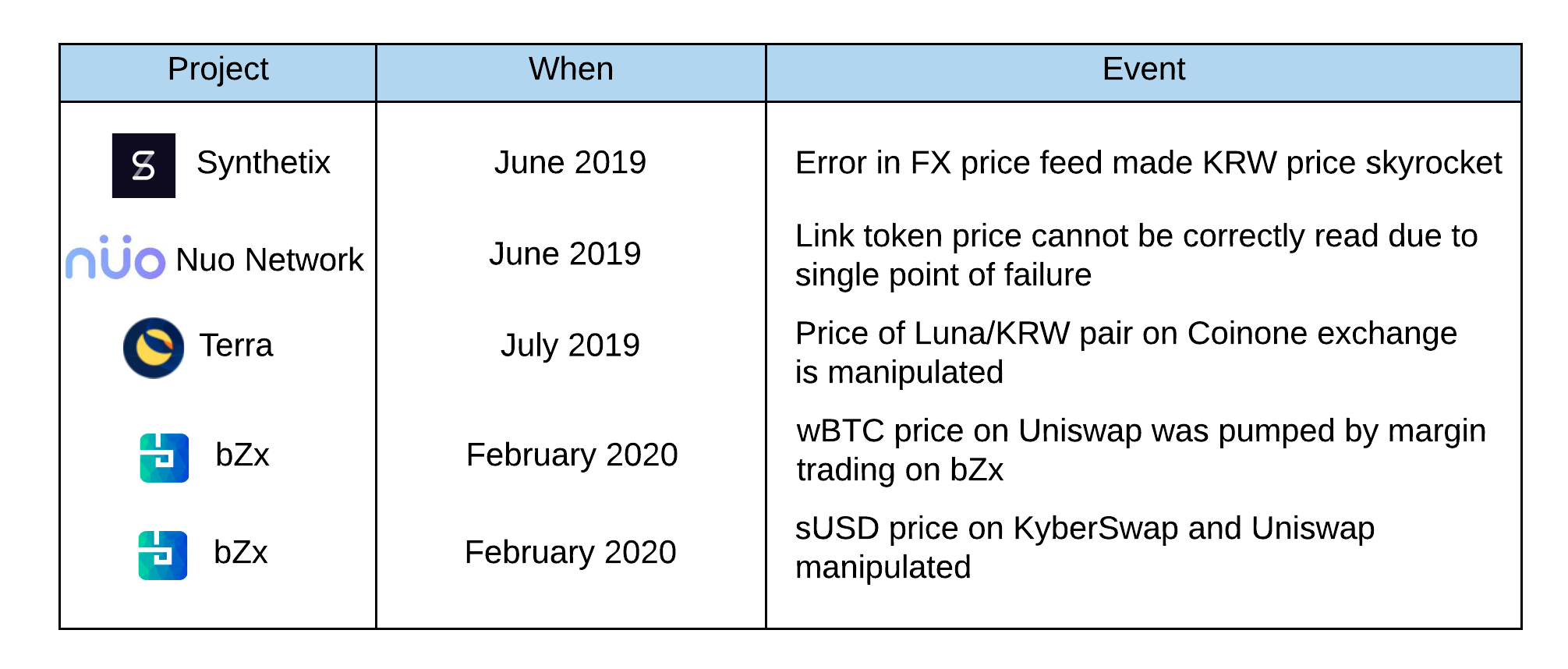}
    \caption{Non-custodial system oracle manipulation events.}
    \label{table:manipulation_events}
\end{table*}

\subsection{Reserve Fund Stablecoins}
\label{sec:appendix-reservefund}

Reserve Fund stablecoins can be modeled as Exchange-Traded Funds (ETFs).\footnote{To account for risk in underlying commercial bank deposits, we can also add a bank run model in serial to an ETF model.} 
In ETFs, an investment vehicle (the ETF) is created with indirect claims to a portfolio of underlying assets (e.g., stocks, bonds, and commodities) held by a custodian.\footnote{ETFs can provide simpler access to underlying portfolio, which may not be accessible to the investor otherwise, and reduced frictions/fees in maintaining small positions.} 
A set of \emph{authorized participants} (APs) are allowed to redeem shares of the ETF for the underlying assets and create new shares of the ETF by depositing underlying assets at the net asset value (NAV). 
The ETF price is pegged to the NAV. 
This peg is maintained by the APs, who capture arbitrage between the ETF shares and the underlying portfolio. 
If direct redemption is allowed in a Reserve Fund stablecoin, then anyone can be an AP.\footnote{Fees may discourage small redemptions, so that large redeemers are de facto APs.} 
Some stablecoins make no promises about future redeemability; in this case, the de facto AP is the issuer itself.

As with ETFs, given sufficiently liquid collateral, the price target is always maintainable within some bounds through these mechanisms.
The tightness of the bounds, however, depend on the liquidity and volatility of the reserve assets. 
For instance, corporate bond ETFs traded at significant deviations from NAV during the financial crisis in 2008 \cite{kaminska2009} and during the SARS-COV-2 market panic in 2020 \cite{aramonte2020}. 
Even US government bonds, which are normally highly liquid, faced high liquidity stress in March 2020 \cite{rennison2020} with corresponding ETFs facing similar NAV-price deviations.

Empirical analysis of ETFs, e.g., \cite{bendavid2018}, suggest that securities with higher ETF ownership are more volatile, which raises concerns about the ETF mechanism.
While ETF membership leads to wider access and so increased trading volume, the relationship with volatility is unclear as the empirical comparison is not controlled. Rather, we would want to compare with a setting in which the underlying portfolio is as easily accessible without the ETF.
An equilibrium model analysis confirms a more nuanced relationship with volatility. \cite{malamud2016} develops a model of endogenous feedback effects in ETFs, in which the liquidity of the underlying portfolio is influenced by the ETF. 
This model shows that ETFs are exposed to different demand shocks than the underlying basket. Even with small deviations, APs that arbitrage through leveraged positions can amplify the differences.\footnote{As stated in \cite{malamud2016}, ``ETFs may be both a blessing and a curse. 
That is introducing new ETFs may lead to a significant amplification of speculative behavior of arbitrageurs, destablize the market, and lead to a spike in volatility; however, at the same time, a ``good'' ETF may actually stabilize the economy, lead to a significant reduction in volatility, and improve the liquidity of the underlying securities.''}

An ETF-like model is developed for Reserve Fund stablecoins in \cite{lyons2020} and interpreted against Tether trading data.
Models such as these are a natural starting point to address the following open questions about Reserve Fund stablecoins:
\begin{itemize}
    \item \textbf{Issuer AP incentives.} Issuers are in a position to prevent competition and decide timing in capturing arbitrage. There is a trade-off between the size of mispricings before APs intervene, and maintaining a stable asset, which affects demand and ultimately assets under management, for which they are awarded deposit interest.
    \item \textbf{Issuer target incentives.} If the peg target is defined at the discretion of the issuer (e.g., not USD or an external index), then the issuer may have incentive to manipulate the target index to its advantage. For instance, if the stablecoin is large enough, changing the target can have a market impact, which may be advantageous to outside positions held by the issuer.
    \item \textbf{Effects on fiat currencies.} Does stablecoin structure affect the ability of government to stabilize currencies? This is a concern of regulators regarding the size of potential stablecoins, like Libra. This effect could be modeled with ETF structure in series with currency models.
    \item \textbf{Effects on crypto markets.} 
    \cite{griffin2019} suggested that stablecoins have been used to manipulate Bitcoin prices. A model of the economic structure in Bitcoin/stablecoin markets (e.g., \cite{lyons2020}) could help determine the direction of causality suggested by the data.
\end{itemize}

Some of these open questions are relevant to the wider ETF literature itself and are not specific to stablecoins. 

\subsection{Fractional Reserve Fund}
\label{sec:appendix-fractionalreservefund}

\paragraph{Bank Fund}

In a Bank Fund stablecoin, the issuer maintains a balance sheet functionally similar to a commercial bank. This balance sheet is based on fractional reserves with deposit obligations tied to stablecoins that are issued. Aside from the fractional reserve, the bank holds other capital assets that are illiquid and earn a yield for the bank. This is a nearly identical model to a normal bank with a few exceptions: (1) the stablecoin bank my not be regulated or audited, (2) the bank my not be government-insured against bank runs, and (3) the bank may be freer to deny redemptions and/or apply redemption fees.

Bank Fund stablecoins can be understood using bank run models in series with ETF models. 
In a bank run, the fractional liquid reserve of the bank is depleted from redemptions, after which the bank defaults as the bank's remaining assets are illiquid and can only be sold quickly at large discounts (a fire sale). 
In a bank run, remaining depositors' lose their money. \cite{diamond1983} shows multiple equilibria to the game played between depositors. 
This includes a bank run equilibrium, in which all depositors scramble to redeem their deposits, triggering the collapse in a self-fulfilling way. 
One approach is the global games setting of \cite{carlsson1993} adapted to bank runs in \cite{rochet2004} and \cite{goldstein2005}. 
In this setting, depositors observe bank fundamentals with noise (e.g., the reserve ratio could be random), and they will choose to rollover (i.e., extend the maturity of) their deposits if their signal is above a threshold. 
\cite{he2012} introduced a staggered debt structure of deposit maturities. 
A point of difference to existing bank run models are the non-negligible network effects among stablecoin holders, much less so than among  traditional bank depositors. 

Bank runs used to happen somewhat regularly. 
To prevent frequent crises of faith, governments issued depositor insurance against bank runs. 
However, Bank Fund stablecoins are unlikely to have such insurance and so remain susceptible to bank runs. 
A key consideration here is that bank runs follow a threshold effect in depositor faith. After a threshold is reached, too many depositors try to redeem, sending the bank's balance sheet into a `death spiral'. 
Below this threshold, however, the coin may be very stable.

As noted above, a Bank Fund stablecoin may be freer to deny redemptions and/or apply redemption fees. 
An event like this triggered a crisis in Tether in Oct. 2018 (see Table~\ref{table:custodial_events}). 
These levers may also be applied strategically to discourage the continuation of bank runs or could be abused to create profitable price discrepancies for the issuer to arbitrage. 
Thus open questions emerge around issuer incentives as in the Reserve Fund.

\paragraph{Money Market Fund}
In a Money Market Fund an underlying portfolio is meant to closely track a target, with some return. 
A traditional Money Market Fund maintains a fixed NAV for redemptions. 
While the underlying assets are usually highly liquid and relatively stable, their market values float and so there is some risk that the fixed NAV is unsustainable. 
This leads to a liquidity risk related to bank runs: shocks to the underlying assets leads money market funds to liquidate assets, which can have the effect of lowering prices further if liquidity is temporarily constrained, which can cause even more liquidations.
Money Market stablecoins can be understood using money market fund models, e.g., \cite{parlatore2016}, in series with ETF models. 
There are many case studies of money market funds breaking the dollar during the 2008 financial crisis. 
In particular, \cite{Kacperczyk13} show that in the presence of high inflows, money market funds had expanded their risk-taking and they suffered runs as a result. 
Some of the proposed forms of Libra closely resemble money market structures.




\subsection{Discussion of Oracles}
Centralized oracles control the risk of outside attack but can lead to perverse incentives for the provider--at some point, manipulating the feeds may be more profitable than providing data honestly. They also introduce single points of failure. Centralized approaches can be made more secure, for instance, through the use of trusted execution environments \cite{zhang2016}. Through such methods, it can be proven that the data feed is an authentic representation of a particular source, but it is still inherently manipulable by the source.

Decentralized oracle approaches exist, but remain an open research question. Existing solutions fall short of a full solution. They rely on Schelling point schemes, in which agents vote on the price feed and are incentivized by slashing if their vote deviates from the consensus. These are problematic because incentives are related to the consensus, which is not objectively verifiable for correctness and can be manipulable through game theoretic attacks.

There are methods to mitigate these risks. For instance, medianizers are typically used to aggregate prices from a number of oracles, half of which must then be incorrect to manipulate the final feed. Some services, such as Chainlink, provide such a medianizer using an incentivized reputation system \cite{chainlink}. The security of such systems also remains an open question.

Other methods attempt to create a price feed inferred from on-chain metrics, which is then objectively verifiable on-chain \cite{klagesmundt_grant2018}. A related method attempts to couple the price of a token to the cost of mining in proof-of-sequential work (e.g., Elasticoin \cite{dong2020} and Meter \cite{meter-io}).\footnote{Though note that as `stablecoins' Elasticoin and Meter are only upper bounded in price without a risk absorption mechanism. Melmint adds a seigniorage shares mechanism atop Elasticoin to absorb risk.} The security of these methods also remains an open question.

Some cryptocurrency-to-cryptocurrency prices can be determined on-chain through decentralized exchanges, given appropriately controlled construction (e.g., to account for limited liquidity and time-averaged over extended time periods to make manipulation more costly). A missing link is still to outside fiat prices, however. Prices in terms of other stablecoins may be used, but this faces the same inherent problem: we then rely on that stablecoin, which may be manipulated or fail, for the data feed.

\subsection{Agents, preferences and attitudes to risk}
\label{appendix:agents-preferences}

\subsubsection{Utility functions}

Provided an agent's preferences satisfy certain properties, an agents' preferences over consumption set $Y$ can be represented by a utility function~\cite{mas1995microeconomic}.
In particular, here we assume that an agents are \emph{mean-variance} maximizers, roughly wanting to maximize the mean and minimize the variance of a portfolio, with preferences over a random variable $X$ can be described as follows:

\begin{equation}
    U(X) = \mu_X - \frac{\rho_A \sigma_{X}^{2}}{2}
\label{eq:mv-maximizer}
\end{equation}

where $X \stackrel{}{\sim} N(\mu_X, \sigma_X)$, with $\mu_X$ denoting the mean of $X$, $\sigma_X$ denoting the variance and $\rho_A$ denoting the coefficient of risk aversion. 
We provide more information on this formulation in \ref{appendix:hyperbolic}.

\subsubsection{Method 1: one risky asset, one riskless asset}

In one simple framework, a \emph{mean-variance} maximizer can invest proportion $\alpha$ of their wealth in a risky asset, and proportion $(1-\alpha)$ in a risk free asset.
From this setup, it is possible to derive, as we do in \ref{appendix:method-one}, that their optimal choice of $\alpha$ is given as follows: 

\begin{equation}
\label{eq:optimal-soln}
\alpha^{*} w = \frac{\mathrm{E}[R] -r}{\rho_A Var(R)}
\end{equation}

where $w$ denotes the agent's wealth, $\mathrm{E}[R]$ and $Var(R)$ the expected return and variance of a risky asset and$r$ denotes the return on a risk-free asset.
From this expression, all that is required to compute $\rho_A$ is knowledge of the five variables in this equation, making it a tractable place to begin with the estimation of agents' preferences.

\subsubsection{Method 2: preferences from portfolio weights}

It is also possible to uses agents' investment history to infer agents' risk-aversion coefficients. 
In particular, \cite{bodnar15} consider an investor who invests into $k$ risky assets and a single riskless asset, basing their investment strategy on an exponential utility function, as above. 
As well as permitting multiple risky assets, in contrast to above, the closed-form solution to the portfolio choice problem provided by the authors is also explicitly multi-period.
We present the details of this approach in \ref{appendix:method-two}.

\subsubsection{A case study of MakerDAO using Method 1}

\begin{figure}
    \centering
    \includegraphics[width=\columnwidth]{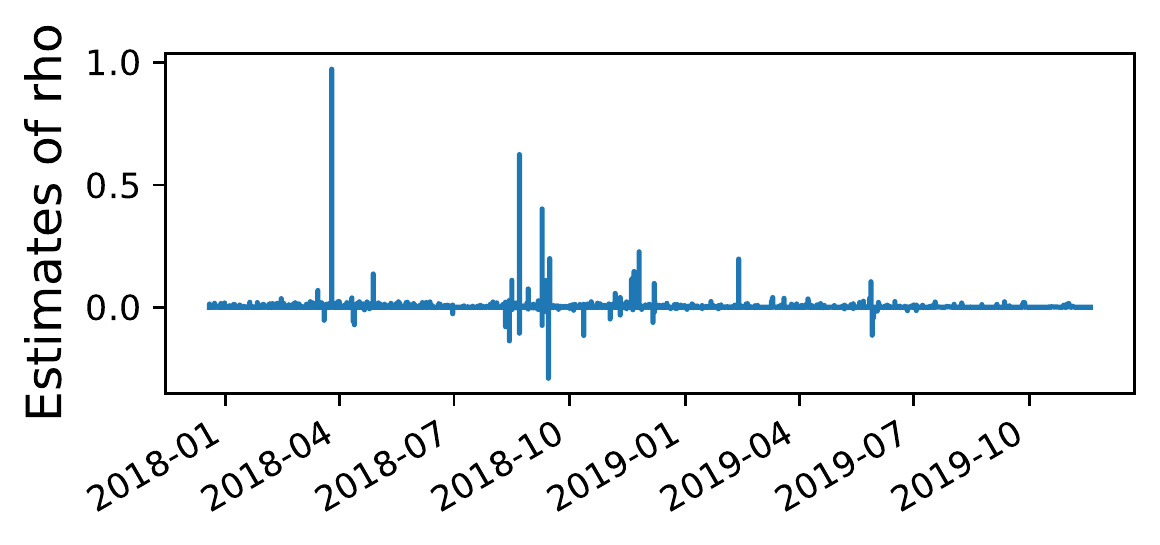}
    \caption{Values of $\rho$ per CDP. }
    \label{fig:rho-cdp-level}
\end{figure}

\begin{figure}
    \centering
    \includegraphics[width=\columnwidth]{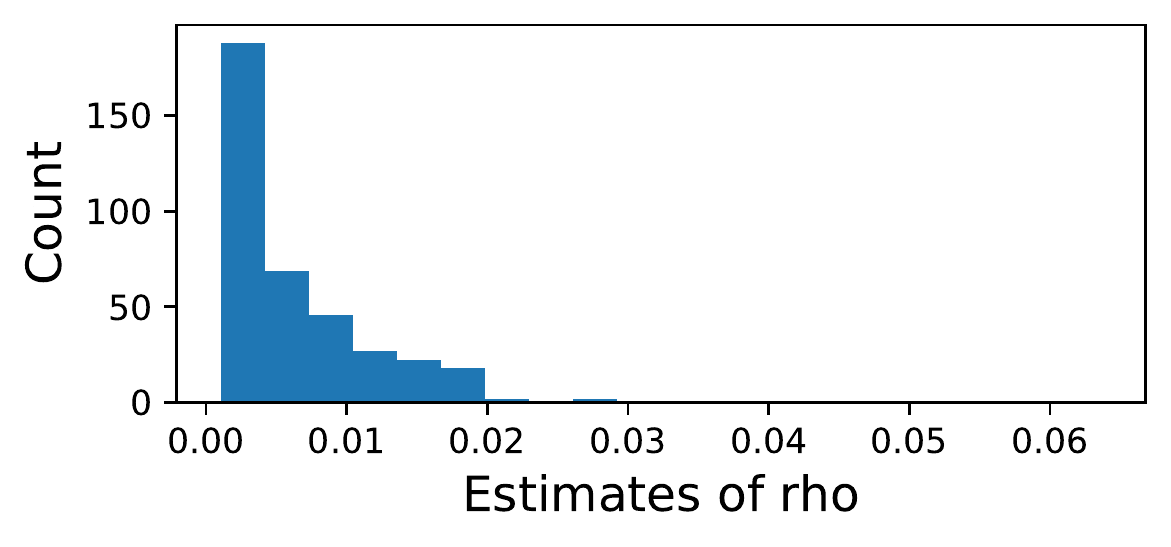}
    \caption{Values of $\rho$ per Externally Owned Account.}
    \label{fig:rho-address-level}
\end{figure}

We apply Method 1 to Equation~\ref{eq:optimal-soln} to seek to recover agents' risk aversion in choosing leverage in the MakerDAO protocol~\cite{makerdao}, a non-custodial collateral backed stablecoin (see Section~\ref{sec:noncustodial}).
We use data on single collateral Dai (Sai) up until November 18th 2019. 
A histogram of the resulting values of $\rho$ per CDP is given in Figure~\ref{fig:rho-cdp-level}\footnote{Note that we exclude outliers in the plot, e.g. those with risk aversion above 1}.
While these results should only be considered indicative, we find a mean value for $\rho$ of 0.0011, which seems approximately consistent with other estimates of risk-aversion coefficients in the literature~\cite{babcock1993risk}.
We also provide an average value of $\rho$ per address, rather than per CDP, in Figure~\ref{fig:rho-address-level}.
Looking at `active' accounts with more than 10 CDP actions, we find a mean value for $\rho_A$ of 0.0012.
The main takeaway from figure~\ref{fig:rho-address-level} is that on an address level, most addresses appear to exhibit some degree of risk aversion, with some estimates of $\rho$ providing notably higher levels of risk aversion than appear in the literature.

\subsubsection{Utility function estimation - details.}
\label{appendix:hyperbolic}

We take as our starting point a general class of utility functions: those representing Hyperbolic Absolute Risk Aversion (HARA), where the level of risk tolerance is a linear function of wealth:

\begin{equation}
\label{eq:HARA}
u(w) = \frac{1-\gamma}{\gamma} \left[ \frac{a w}{1-\gamma} + b \right]^\gamma
\end{equation}

where $u(w)$ is the utility arising form a certain level of wealth $w$,
$a>0$, $\gamma \neq 0$ and $\frac{a w}{1-\gamma} + b > 0$.
A standard measure of risk is the Arrow-Pratt coefficient of absolute risk-aversion~\cite{arrow1965aspects, pratt1978risk}, which extracts a measure of risk-aversion that is invariant to affine transformations as follows: \footnote{See~\cite{mas1995microeconomic} for further information on expected utility theory and the relevance of affine transformations.}

\begin{equation}
\label{eq:arrow-pratt-absolute}
A(w) = - \frac{u''(w)}{u'(w)}
\end{equation}

Importantly, imposing parameter restrictions $a>0$, $b = 1$ and $\gamma \rightarrow - \infty$ (~\cite{sengupta2016decision} on equation (\ref{eq:HARA}) yields an exponential utility function $u(w) = -e^{-a w}$, with the property of \textit{constant absolute risk aversion (CARA)}: $A(w) = - \frac{- a ^ 2 e^{-a w}}{a e^{-a w}} = a = \rho_A$.
CARA implies that the amount an agent optimally invests in a risky asset does not depend on their wealth. 
In turn, assuming that agents' utility functions feature can be characterized as CARA,
then for random variable $X$, provided $X \stackrel{}{\sim} N(\mu_X, \sigma_X)$ where $\mu_X$ denotes the mean of $X$ and $\sigma_X$ denotes the variance, it can be shown that the expected utility $\mathrm{E}[u(X)]$ is given by $\mathrm{E}[u(X)] = - e ^ {-\rho_A \left[ \mu_X - \frac{\rho_A \sigma_{X}^{2}}{2} \right]}$~\cite{thomas1987macroeconomic}.
The agent maximizes this expected utility when they maximize $\mu_X - \frac{\rho_A \sigma_{X}^{2}}{2}$.
Therefore, if we characterize an agent as having exponential utility, and therefore CARA, then when they maximize this utility when faced with a normally distributed random variable $X$, they can be considered a \emph{mean-variance maximizer}, with utility given by:

\begin{equation}
    U(X) = \mu_X - \frac{\rho_A \sigma_{X}^{2}}{2}
\end{equation}

Treating agents as mean-variance maximizers yields one tractable framework within which agents risk aversion, an aspect of their preferences, can be measured. 
Yet there are several points to note about this approach. 
Firstly, assuming that agents exhibit CARA---where their investment in a risky asset does not depend on their wealth---may not be wholly realistic. 
Perhaps agents actually invest a constant \textit{proportion} of their wealth.
Moreover, here we are implicitly assuming that agents are not concerned with the shape of the risk, aside from the variance, so for instance are not concerned with heavy tails. 
In the stablecoin setting, this may too be an unrealistic representation of the true distributions. 
We note these limitations and posit this framework as a tractable entry point for future research. 


\subsubsection{Method 1: one risky asset, one riskless asset}
\label{appendix:method-one}

Let us assume that an agent can invest proportion $\alpha$ of their wealth in a risky asset, and proportion $(1-\alpha)$ in a risk free asset.\footnote{Here we are not considering the participation question about whether to invest at all, but instead considering how, given a fixed amount to invest, this can be done optimally.}
This would provide a total return $X(\alpha) = \alpha R + (1-\alpha)r$.
Since $\mathrm{E}[X(\alpha)] = r + \alpha (\mathrm{E}[R] - r)$ and $var(X(\alpha)) = \alpha ^ 2 var (R)$, setting $\mu_X =\mathrm{E}[X(\alpha)]$ and $\sigma_X ^2 =var(X(\alpha))$, an agent with wealth $w$ will maximize

\begin{equation}
\label{eq:agent-max-pb}
w[r + \alpha(\mathrm{E}[R] -r)] - \frac{1}{2}\rho_A w^2 \alpha ^2 Var(R)
\end{equation}

with respect to $\alpha$, yielding optimal solution as given in Equation~\ref{eq:optimal-soln}.
From Equation~\ref{eq:optimal-soln} all that is required to compute $\rho_A$ is knowledge of the five variables in this equation, making it a tractable place to begin with the estimation of agents' preferences.

\subsubsection{Method two: preferences from portfolio weights}
\label{appendix:method-two}

Letting $\mathbf{X}_\tau$ be a random return vector of $k$ risky assets, and supposing that $\mathbf{X}_\tau$ and a vector of $p$ predictable variables $\mathbf{z}_\tau$ jointly follow a vector autoregressive process of order 1, the authors prove that the optimal multi-period portfolio weights for all periods $[0, T-1]$ can be analytically stated. 
In particular, by Corollary 2, letting $\mathbf{X}_{\tau} = (X_{\tau , 1},X_{\tau , 2},...,X_{\tau , k})'$ be a sequence of independently and identically normally distributed vectors of $k$ risky assets ($\mathbf{X}_{\tau} \stackrel{}{\sim} N(\mathbf{\mu}, \mathbf{\Sigma})$), $r_{f, \tau}$ be the riskless asset return, and provided $\Sigma$ is positive definite, then $\forall t=1, ... T$:

\begin{equation}
\label{eq:bodnar-equation}
\mathbf{w_{T-t}^{*} = \frac{1}{\rho_A W_{T-t} \Pi_{i=T-t+2}^{T} R_{f,i}} \mathbf{\Sigma}^{-1} \mathbf{\hat{\mu}}}
\end{equation}

where $\hat{\mu} = \mu - r_{f, T-t+2} \mathbf{1}$, which can be rearranged to yield an explicit expression for $\rho_A$:

\begin{equation}
\label{eq:bodnar-equation-rearranged}
\rho_A  = \frac{1}{\mathbf{w_{T-t}^{*} W_{T-t} \Pi_{i=T-t+2}^{T} R_{f,i}} }\mathbf{\Sigma}^{-1} \mathbf{\hat{\mu}}
\end{equation}

On this approach, provided data is available on agents' portfolio weights through time, a value for $\rho_A$ could potentially be calibrated more precisely than method one would allow; however, this data requirement in itself is more demanding. 
In particular, in the context of stablecoins, for example, the possibility that one agent uses multiple blockchain addresses would obfuscate the true portfolio weights through time. 
However, to the extent that future work is able to accurately determine these weights, this offers a promising approach to calibrate values of $\rho_A$.

\subsubsection{Empirical case study of Method 1}
\label{appendix:method-one-empirical}
To illustrate how these utility function estimation techniques can be applied, we provide a minimal working example, applying method 1 to MakerDAO~\cite{makerdao}.

A core component of the Maker stablecoin system is the issuance of a stablecoin against the value of collateral.
In particular, down to a threshold value of 150\%, agents choose how much stablecoin to issue as debt against their collateral. 
For example, for 150 USD worth of ETH collateral, at the 150\% threshold an agent can issue up to 100 USD of stablecoin debt.
However, if the ETH/USD price falls, then the agent would become undercollateralized relative to the 150\% threshold, and would incur liquidation costs. 
On the converse---and one of the primary use cases of such a stablecoin---if the agent repurchases more ETH with their debt, the agent has accessed leverage. 
If the ETH/USD price rises, then the agent will stand to benefit more from this price increase than if they had not issued themselves debt. 

Thus, following method 1, in this section the goal is to estimate equation (\ref{eq:optimal-soln}).
We proceed with the following demonstrative steps.

\begin{enumerate}
    \item \textbf{Data collection.} We use the MakerDAO GraphQL API~\cite{makergraphql} to obtain data on Collateralized Debt Position (CDP) actions.\footnote{This API only covers the stablecoin SAI, the precursor to DAI.}
    \item \textbf{Data cleaning and sample selection.} We clean the data, focusing only on Externally Owned Accounts prior to the launch of multi-collateral DAI. We further only consider CDPs with more than 50 USD of collateral.
    \item \textbf{Wealth calculation ($w$).} We assume that each time an agent issues themselves with the stablecoin, this is used to buy more ETH. Therefore for each agent we calculate their total wealth as the sum of their ETH holdings (ETH collateral and ETH bought with stablecoin) less their debt.
    \item\textbf{Risky asset holding ($\alpha$).} We calculate the ratio of ETH holdings to original ETH collateral. Leverage is represented as $\alpha > 1$.
    \item \textbf{Computation of mean and variance of risky asset ($\mathrm{E}[R]$ and $Var(R)$).} We compute the mean and variance of the risky asset by computing the cumulative rolling moving average mean and variance of daily ETH/USD returns.
    \item \textbf{Assumption of a risk free rate ($r$).} We assume that the investor has access to a risk-free interest rate of 2\% annually. 
\end{enumerate}

\subsection{Cross-chain and Synthetic Assets}
\label{sec:appendix-crosschain}

Synthetic assets use the same mechanisms as non-custodial stablecoins but with different target pegs (e.g., dYdX's perpetuals using synthetic BTC).
In comparison, cross-chain mechanisms transfer assets between blockchains.
Where both blockchains are able to verify state of the other, cross-chain assets do not require collateral as the issue and redeem procedures can be executed through transaction inclusion proofs via a chain relay on each blockchain (e.g. PeaceRelay~\cite{Luu17PeaceRelay}).
Hence, incentive design for cross-chain mechanisms is not required to maintain a price peg, but rather to keep the relays on each side up-to-date and protected against attacks such as relay poisoning~\cite{Zamyatin2019XCLAIM, Interlay20StakedRelayers}.

If a cross-chain mechanism enables asset transfers (i.e., not atomic swaps) from a blockchain which does \textit{not} have the ability to verify the state of another blockchain (e.g., Bitcoin) to one that does (e.g., Ethereum), collateral or trust in a third party is required.\footnote{For a formal proof of this requirement see~\cite{Zamyatin2020SoK}.}
These cross-chain mechanisms utilize intermediaries that hold custody over the locked asset.
We can distinguish between trusted non-collateralized intermediaries where custodial models can be applied (e.g., wBTC) and non-custodial cross-chain mechanisms (e.g., XCLAIM, tBTC, RenBTC).
Non-custodial designs rely on collateral for incentive security in addition to collateral of the transferred asset itself.

Exogenous collateral without governance assets (e.g. XCLAIM~\cite{Zamyatin2019XCLAIM, Harz2019}) can be modelled using the capital structure models without considering the long-term impact of governance token value.
Models that use exogenous collateral for the transferred asset in combination with endogenous collateral for incentives (e.g. tBTC), might be subject to a similar governance token value problem as outlined in~\ref{sec:incentive-sustain}.
However, in both cases the underlying asset is insured by exogenous collateral and hence the design provides protection of the transferred assets independent of the success of the cross-chain mechanism.
Endogenous collateral structures, on the other hand, are subject to the same incentive sustainability issues that rely on an increasing governance token value (e.g. RenBTC).
Here, the security of the transferred asset relies on the long-term success of the cross-chain mechanism to disincentivize attacks.

\end{document}